\documentclass[lettersize,journal]{IEEEtran}
\usepackage{amsmath,amsfonts}
\usepackage{algorithmic}
\usepackage{algorithm}
\usepackage{array}
\usepackage[caption=false,font=footnotesize]{subfig}
\usepackage{textcomp}
\usepackage{stfloats}
\usepackage{url}
\usepackage{verbatim}
\usepackage{graphicx}
\usepackage{cite}
\usepackage{hyperref}

\usepackage{soul}

\usepackage{multirow}
\usepackage{xcolor}
\usepackage{cuted}
\usepackage{amsthm}
\usepackage{mathtools}

\newtheorem{proposition}{Proposition}
\newtheorem{corollary}{Corollary}
\newtheorem{lemma}{Lemma}
\newtheorem{assumption}{Assumption}
\newtheorem{example}{Example}

\usepackage{amsmath,amsfonts,bm}

\def\eqref#1{equation~\ref{#1}}

\def\1{\bm{1}}

\def\va{{\bm{a}}}

\def\vd{{\bm{d}}}

\def\vh{{\bm{h}}}

\def\vs{{\bm{s}}}

\def\vu{{\bm{u}}}
\def\vv{{\bm{v}}}

\def\vz{{\bm{z}}}

\def\mC{{\bm{C}}}

\def\mE{{\bm{E}}}

\def\mH{{\bm{H}}}
\def\mI{{\bm{I}}}

\def\mP{{\bm{P}}}

\def\mR{{\bm{R}}}

\def\mT{{\bm{T}}}
\def\mU{{\bm{U}}}
\def\mV{{\bm{V}}}

\DeclareMathAlphabet{\mathsfit}{\encodingdefault}{\sfdefault}{m}{sl}
\SetMathAlphabet{\mathsfit}{bold}{\encodingdefault}{\sfdefault}{bx}{n}

\def\sC{{\mathbb{C}}}

\newcommand{\sGamma}{\widehat{\mathbf{\Gamma}}}
\newcommand{\mGamma}{\mathbf{\Gamma}}

\DeclareMathOperator*{\argmax}{arg\,max}
\DeclareMathOperator*{\argmin}{arg\,min}

\usepackage{xr}
\makeatletter

\newcommand*{\addFileDependency}[1]{
\typeout{(#1)}
\@addtofilelist{#1}
\IfFileExists{#1}{}{\typeout{No file #1.}}
}\makeatother

\begin{document}

\title{On Interference-Rejection Using Riemannian Geometry for Direction of Arrival Estimation}

\author{Amitay Bar and Ronen Talmon 
\thanks{A. Bar and R. Talmon are with the Viterbi Faculty of Electrical and Computer Engineering, Technion—Israel Institute of Technology, Haifa 32000, Israel (e-mail: amitayb@campus.technion.ac.il; ronen@ef.technion.ac.il). 

This work was supported by the European Union’s Horizon 2020 research and innovation programme under grant agreement No. 802735-ERC-DIFFOP.
}
}



\maketitle

\begin{abstract}
We consider the problem of estimating the direction of arrival of desired acoustic sources in the presence of multiple acoustic interference sources. All the sources are located in noisy and reverberant environments and are received by a microphone array. We propose a new approach for designing beamformers {and DoA estimation methods} based on the Riemannian geometry of the manifold of Hermitian positive definite matrices. Specifically, we show theoretically that incorporating the Riemannian mean of the spatial correlation matrices into frequently-used beamformers gives rise to {spatial spectra} that reject the directions of interference sources and result in a higher signal-to-interference ratio. We experimentally demonstrate the advantages of our approach in designing several beamformers {and a recent DoA estimation method} in the presence of simultaneously active multiple interference sources.
\end{abstract}


\section{Introduction}
\IEEEPARstart{E}{stimation} of the direction of arrival (DoA) of an acoustic source is prevalent in signal processing; it is an important step in many tasks, such as source localization, beamforming, source separation, spectrum sensing, and speech enhancement \cite{gannot2019introduction}, to name but a few. 
Despite the large research attention it has drawn in the past decades, acoustic DoA estimation is still considered a challenging open problem. Especially in noisy and reverberant environments and in the presence of interference sources, it continues to be an active research field. 

Acoustic source localization, and particularly DoA estimation, are often addressed using beamforming \cite{krishnaveni2013beamforming}.
Many beamformers have been proposed over the years for these tasks. One class of beamformers is based on the steered response power (SRP) of a beamformer output. For example, considering the maximum likelihood criterion for a single source, the output power of the beamformer from all the directions is computed, and the DoA is identified as the direction with the maximal power \cite{widrow1967adaptive,compton1988adaptive,vook1992bandwidth, yao2002maximum}. 
Another example is the Minimum Variance Distortionless Response (MVDR) beamformer \cite{akbari2010music,salvati2016weighted,rieken2004generalizing}, which was first introduced by Capon \cite{capon1969high}. The MVDR beamformer extracts the DoA of each of the existing sources, maintaining a unit gain at their direction while minimizing the response from other directions.
An important generalization of the MVDR beamformer is the Linearly Constrained Minimum Variance (LCMV) beamformer \cite{frost1972algorithm}, obtained by minimizing the output power under multiple linear constraints, and can be used for DoA estimation as well \cite{xu2015response}.
Another line of beamformers is derived based on a subspace approach, i.e., by identifying the subspace of the desired sources, which is assumed to contain only a small portion of the noise and the interference sources. A prominent subspace method, which is also used for DoA estimation, is MUltiple Signal Classification (MUSIC) \cite{schmidt1986multiple,yan2013low,zhang2010direction,vallet2015performance}.

{A notable algorithm for acoustic source localization is the Steered-Response Power Phase Transform (SRP-PHAT) algorithm proposed in \cite{dibiase2001robust}. In SRP-PHAT, the phase transform is used to normalize the different frequencies, such that only their phase information is considered. This allows for the fusion of the different frequencies when considering a broadband signal. A popular time-domain implementation of the SRP-PHAT is the generalized cross-correlation with phase transform (GCC-PHAT) proposed in \cite{knapp1976generalized}, which normalizes each cross-correlation using the phase transform.}
{In recent years, sparse signal recovery methods have been proposed for DoA estimation \cite{malioutov2005sparse,hu2013doa,tan2013sparse}. In particular, one approach for solving the sparse recovery problem is the sparse Bayesian learning approach proposed in \cite{tipping2001sparse}, which was adopted for the DoA estimation problem as well, for example in  \cite{hu2016source,dai2021real}. }
    
In this paper, we consider DoA estimation in a reverberant enclosure consisting of desired sources along with interference sources. We assume that the desired sources are constantly active, whereas the interference sources are only intermittently active. 
The number of sources, their locations, and their times of activity are all unknown. Consequently, their identification as desired or interference is unknown as well. The power of the different sources is also unknown, and the interference sources could, in fact, be stronger than the desired sources with overlapping activity periods.
Our goal is to estimate the DoA of the desired sources in the presence of possibly simultaneously active, multiple interference sources. 

This setting poses a major challenge to the common practice in existing methods that rely on maximal power because estimating the DoA of the strongest sources might result in distinct beams in the direction of the interference sources rather than the desired sources. Furthermore, these beams could mask the beams pointing at the directions of desired sources.

We address this challenge from a geometric standpoint. Our approach relies on the observation that the frequently-used beamformers implicitly consider Euclidean geometry when processing sample correlation matrices. Therefore, since the sample correlation matrices are Hermitian Positive Definite (HPD) matrices, important geometric information is not fully utilized.
Instead, we propose a new approach for beamforming design that is based on the Riemannian geometry of the manifold of HPD matrices \cite{bhatia2006riemannian,nielsen2013matrix,skovgaard1984riemannian}. 
Concretely, we analyze the received signal in short time windows and consider the Riemannian mean \cite{bhatia2006riemannian} of the sample correlation matrices in these windows. Then, we leverage particular spectral properties of the Riemannian mean. In \cite{katz2020spectral}, it was shown that the Riemannian mean of HPD matrices preserves shared spectral components and attenuates unshared spectral components.
Consequently, the continual activity of the desired sources and the intermittent activity of the interference sources enable us to associate desired sources with shared spectral components and interference sources with unshared spectral components. By combining the above, we show that the incorporation of the Riemannian mean into the beamformer design leads to interference rejection, i.e.,  gives rise to {a spatial spectrum} that implicitly rejects the beams pointing at the interference sources and preserves the beams pointing at the desired sources. The resulting {spectrum} is, in turn, used for the estimation of the DoA of the desired sources.
Importantly, our approach is applicable to a large number of beamformers used for DoA estimation.

By incorporating our Riemannian approach, we present new implementations of several beamformers: the Delay and Sum (DS) beamformer, subspace-based beamformers, and the MVDR beamformer, {as well as the Bayesian learning method for DoA estimation proposed in \cite{hu2016source}.}

{The main contributions of this paper are as follows. 
First, we observe that despite being commonly-used in array processing in general, and for DoA estimation in particular, the HPD structure of the signal correlation matrices is not fully exploited. %
Here, we identify that by dividing the computation of the sample correlation matrix into short-time segments, it can be viewed as the Euclidean mean of the sample correlation matrices of the segments. This observation enables the introduction of the Riemannian geometry of HPD matrices, and in particular, it allows us to introduce a rather simple approach that promotes the use of the Riemannian mean instead of the Euclidean mean and to show its multiple merits. 
Second, we demonstrate that the proposed approach can naturally be applied to a broad range of DoA estimation methods that use the signal sample correlation matrix, and we show that it is computationally efficient in the sense that it typically does not change the order of the computational complexity of the DoA estimation methods.
Third, we theoretically analyze the proposed approach and show that in the case of the DS beamformer, it results in higher SIR values that lead to better DoA estimation accuracy. In addition, we present a noise sensitivity analysis.
Fourth, we present empirical results in adverse conditions that include simultaneously active multiple interference sources. We showcase the applicability of our approach to both classical and recent DoA estimation methods and demonstrate that the obtained performance improvement is by a large margin. }

We conclude the introduction with three remarks.
First, a similar setting to ours, consisting of desired sources accompanied by interference sources, was considered in \cite{habets2011speech} and \cite{markovich2009multichannel} but in the context of signal enhancement. In \cite{habets2011speech}, a single desired source and a single interference source were considered, and in \cite{markovich2009multichannel}, multiple desired sources and multiple interference sources were considered. However, in both works, it was assumed that there is at least one segment for each source, desired or interference, in which it is the only active source. 
Furthermore, in \cite{markovich2009multichannel}, the number of the desired sources and their activity patterns were assumed to be known. 
Second, in the context of radar, the Riemannian geometry of the Toeplitz HPD matrices was used in \cite{arnaudon2013riemannian} and \cite{chahrour2021target}
for target detection by comparing Riemannian distances to a threshold. In the radar settings, \cite{chahrour2020improved} estimated the correlation matrix as a linear combination of correlation matrices, with weights that are based on the Riemannian distance.
Third, in this paper, we demonstrate the Riemannian approach for designing beamformers that reject interference sources for DoA estimation. However, other applications, e.g., signal enhancement, could also benefit from {spectra} that reject interference sources. 

This paper is organized as follows. In section \ref{sec: Background on the HPD Manifold}, we present a brief background on the HPD manifold. In Section \ref{sec: Problem Formulation}, we formulate the problem and the setting. In Section \ref{Proposed Approach} we describe the proposed approach and present the algorithm for DoA estimation. In Section \ref{sec: Analysis}, we provide a theoretical analysis of the proposed approach for the DS beamformer. In Section \ref{sec: Extension to other Beamformers}, extensions of the approach to other beamformers are presented. Section \ref{sec: simulation results} shows simulation results demonstrating our Riemannian approach. Lastly, we conclude the work in Section \ref{sec: Conclusion}.

\section{Background on the HPD Manifold}
\label{sec: Background on the HPD Manifold}
An HPD matrix, $\mathbf{\Gamma}\in\sC^{n\times n}$, is a Hermitian matrix, i.e. $\mathbf{\Gamma} = \mathbf{\Gamma}^H$, where $\mathbf{\Gamma}^H$ is the conjugate transpose of $\mathbf{\Gamma}$, whose real eigenvalues are strictly positive. Associating the space of HPD matrices with the Affine Invariant metric \cite{hiai2009riemannian} constitutes a Riemannian manifold, $\mathcal{M}$. The distance between two matrices $\mathbf{\Gamma}_1$ and $\mathbf{\Gamma}_2$, induced by the Affine Invariant metric, is given by
\begin{equation}
    d^2_{\text{R}}(\mathbf{\Gamma}_1,\mathbf{\Gamma}_2) = 
    \Big\|\log \left( \mathbf{\Gamma}_2^{-\frac{1}{2}}\mathbf{\Gamma}_1 \mathbf{\Gamma}_2^{-\frac{1}{2}} \right)\Big\|_{F}^2,
\end{equation}
where $\|\cdot\|_{F}$ is the Frobenius norm.

{The tangent space to a point on a manifold $\mathbf{\Gamma}\in \mathcal{M}$ is a vector space that can be viewed as a local approximation of the manifold around $\mathbf{\Gamma}$. For the HPD manifold, the tangent space around each $\mathbf{\Gamma}\in \mathcal{M}$ is the vector space of Hermitian matrices of dimensions $n\times n$.  %
} 

The Riemannian mean, $\mathbf{\Gamma}_{\text{R}}$, of a set of point $\{\mathbf{\Gamma}_i | \mathbf{\Gamma}_i\in\mathcal{M}\}$ is defined by the Fr\'echet mean {as the point minimizing the sum of the distances from all the points in the set} as follows
\begin{equation}
    \mathbf{\Gamma}_{\text{R}} \equiv
    \argmin_{\mathbf{\Gamma}\in\mathcal{M}} \sum_i d^2_{\text{R}}(\mathbf{\Gamma},\mathbf{\Gamma}_i).
\end{equation}
In general, there is no closed-form expression for the Riemannian mean of more than two matrices, and a solution can be found using an iterative procedure \cite{moakher2005differential}, {described in Algorithm \ref{Alg: Riemannian mean interations}.}
\begin{algorithm}
\caption{Riemannian mean for the HPD manifold \cite{barachant2013classification}}
\label{Alg: Riemannian mean interations}
\textbf{Input:} a set of $K$ HPD matrices $\{\mathbf{\Gamma}_j\}_{j=1}^K$ \\
\textbf{Output:} the Riemannian mean $\mathbf{\Gamma}_{\text{R}}$ \\
\begin{algorithmic}[1]
\STATE Compute ${\mathbf{\Gamma}_{\text{R}}} = \frac{1}{K}\sum_{j=1}^K\mathbf{\Gamma}_j$
\STATE \textbf{do}
\begin{enumerate}
    \item Compute the Euclidean mean in the tangent plane: 
        $\overline{P} =
        \frac{1}{K}\sum_{j=1}^K\text{Log}_{{\mathbf{\Gamma}_\text{R}}}(\mathbf{\Gamma}_j$)
    \item Update ${\mathbf{\Gamma}_{\text{R}}} = \text{Exp}_{\mathbf{\Gamma}_\text{R}}(\overline{P})$
    \item Stop if $\|\overline{P}\|_{\text{F}} < \epsilon$ ($\|\cdot\|_\text{F}$ is the Frobenius norm)
\end{enumerate}
\end{algorithmic}
\end{algorithm}
The computation of the Riemannian mean requires two maps. The Logarithm map, which {maps} an HPD matrix $\mathbf{\Gamma}_i\in \mathcal{M}$ to the tangent space of the HPD manifold at $\mathbf{\Gamma}$, is given by
\begin{equation}
    \text{Log}_{\mathbf{\Gamma}}(\mathbf{\Gamma}_i) = 
    \mathbf{\Gamma}^{\frac{1}{2}}\log(\mathbf{\Gamma}^{-\frac{1}{2}}\mathbf{\Gamma}_i\mathbf{\Gamma}^{-\frac{1}{2}})\mathbf{\Gamma}^{\frac{1}{2}}.
\end{equation}
The Exponential map, which {maps} a vector $\mT$ from the tangent space at $\mathbf{\Gamma}$, is given by
\begin{equation}
    \text{Exp}_{\mathbf{\Gamma}}(\mT) = 
    \mathbf{\Gamma}^{\frac{1}{2}}\exp(\mathbf{\Gamma}^{-\frac{1}{2}}\mT\mathbf{\Gamma}^{-\frac{1}{2}})\mathbf{\Gamma}^{\frac{1}{2}}.
\end{equation}
{We note that there exists an efficient iterative estimator for the Riemannian mean \cite{ho2013recursive}. It is described in Appendix \ref{subsec: Extension to a Streaming Setting}}.
A discussion about the choice of the Affine Invariant metric appears in Appendix \ref{Appendix: On the Particular Choice of the Riemannian Metric}.

\section{Problem Formulation} 
\label{sec: Problem Formulation}
We consider the problem of localizing $N_{\text{D}}$ desired sources in the presence of $N_{\text{I}}$ interference sources. %
All the sources are static and located in a reverberant environment. The signals are received at a noisy microphone array of $M$ microphones, which are positioned in known, but possibly arbitrary, positions.
The acoustic environment between each source and each microphone is modeled by the Acoustic Impulse Response (AIR).
The signal at the $m$th microphone, {considering a far-field setting,} is given by:
\begin{multline}
    z_m(n) =
    \sum_{j=1}^{N_{\text{D}}}s_j^d(n)*h_{jm}^d(n) 
    + \sum_{j=1}^{N_{\text{I}}}s_j^i(n)*h_{jm}^i(n) \\
    +  v_m(n),
\end{multline}
where $s_j^d(n)$ is the $j$th desired source, $s_j^i(n)$ is the $j$th interference source, and $v_m(n)$ is the $m$th microphone noise. %
We denote by $h_{jm}^d(n)$ the AIR between the $m$th microphone and the $j$th desired source. Similarly, we denote by $h_{jm}^i(n)$ the AIR between the $m$th microphone and the $j$th interference source. 

The sources are characterized by their activation times. The desired sources are active during the entire interval.  In contrast, the interference sources are only partially active, namely active during segments of the interval.  
  {We emphasize that albeit we consider intermittent interference sources, longer signals might consist of a larger number of interference sources in addition to more occurrences of active interference sources. We do not assume there is a segment with only the desired source. Longer intervals do not necessarily increase the probability of such a segment since new interference sources could emerge in such a scenario.
    Consequently, even a long signal still poses a significant challenge to the DoA estimation.} 
Additionally, we assume that the desired sources, the interference sources, and the noise are all uncorrelated. The noise is assumed to be spatially white. 
{We note that these assumptions are made for simplicity. Empirically, the proposed approach leads to superior results for correlated signals and for noise with arbitrary covariance matrix as well. }

The received signal is processed using the Short-Time Fourier Transform (STFT).
We denote by $s_j^d(l,k)$ the STFT at the $l$th window and the $k$th frequency of $s_j^d(n)$. The notation for $s_j^i(l,k)$, $h_{jm}^d(l,k)$, $h_{jm}^i(l,k)$, and $v_m(l,k)$ follows similarly.
Then, the STFT at the $l$th window and the $k$th frequency of the received signal is given by
\begin{multline}
        z_m(l,k) = 
        \sum_{j=1}^{N_{\text{D}}}s_j^d(l,k)h_{jm}^d(l,k) + 
        \sum_{j=1}^{N_{\text{I}}}s_j^i(l,k)h_{jm}^i(l,k) \\
        +  v_m(l,k),
\end{multline}
where we assume the length of the window is much larger than the AIR length. %

We stack the received signals, $\{z_m(l,k)\}_m$, from all the microphones to obtain a column vector $\vz(l,k)\in\sC^{M\times 1}$
\begin{equation}
    \begin{split}
        \vz(l,k) =
        [z_1(l,k) \;\; ... \;\; z_M(l,k)]^\top.
    \end{split}
\end{equation}
Its explicit expression is 
\begin{equation}
    \label{eq: z(l,k)}
    \begin{split}
        \vz(l,k) = 
        \mH^d(l,k)\vs^d(l,k) + 
        \mH^i(l,k)\vs^i(l,k) +
        \vv(l,k),
    \end{split}
\end{equation}
where $\vs^d(l,k)$ and $\vs^i(l,k)$ denote the stacked STFT representations of the desired sources and the interference sources, respectively, and are given by
\begin{equation}
    \begin{split}
        \vs^d(l,k) & = 
        [s^d_1(l,k) \;\; ... \;\; s^d_{N_\text{D}}(l,k)]^\top \\
        \vs^i(l,k) & = 
        [s^i_1(l,k) \;\; ... \;\; s^i_{N_\text{I}}(l,k)]^\top,
    \end{split}
\end{equation}
and the noise term is
\begin{equation}
    \begin{split}
        \vv(l,k) =
        [v_1(l,k) \;\; ... \;\; v_M(l,k)]^\top.
    \end{split}
\end{equation}
The Acoustic Transfer Functions (ATFs) from the $j$th desired source and the $j$th interference source to the microphone array are
\begin{equation}
\label{eq: notation definitions}
    \begin{split}
        \vh_j^d(l,k) & =
        [h_{j1}^d(l,k) \;\; ... \;\; h_{jM}^d(l,k)]^\top \;\;\;\; j=1,...,N_{\text{D}} \\
        \vh_j^i(l,k) & =
        [h_{j1}^i(l,k) \;\; ... \;\; h_{jM}^i(l,k)]^\top \;\;\;\; j=1,...,N_{\text{I}},
    \end{split}
\end{equation}
and in a matrix form
\begin{equation}
    \begin{split}
        \mH^d(l,k)  & =
        [\vh_{1}^d(l,k) \;\; ... \;\; \vh_{{N}_\text{D}}^d(l,k)]  \\
        \mH^i(l,k)  & =
        [\vh_{1}^i(l,k) \;\; ... \;\; \vh_{{N}_\text{I}}^i(l,k)].  \\
    \end{split}
\end{equation}

Henceforth, we focus on a single frequency bin and omit the frequency index.
Throughout the paper, we refer to $\vz(l)$ as the received signal.
Since all the sources are static the ATFs do not change over time, so in the following, we omit their STFT window index $l$.

{In this work, we focus on a single frequency to demonstrate the Riemannian approach, which is suitable for narrowband signals.
For broadband signals, this is the formulation of only a single frequency. We note that the fusion of the different frequencies is of great importance, and normalization techniques for the different beamformers have been proposed \cite{salvati2014incoherent} and can be applied after our method. }

Our goal is to estimate the direction to the desired sources, given $\vz(l), \;\; l=1,\ldots,L_{\text{STFT}}$, where $L_{\text{STFT}}$ is the number of STFT windows. The main challenge is the existence of interference sources, positioned at unknown locations with possibly high signal power.

\section{Proposed Approach}
\label{Proposed Approach}

Typically, the DoA estimation of a desired source is based on the output of a beamformer.
In this section, we present the proposed approach applied to the Delay-and-Sum (DS) beamformer. In Section \ref{sec: Extension to other Beamformers}, 
we extend the proposed approach to other {DoA estimation methods}.

We consider arbitrary indexing of the microphones in the array and designate the first microphone as the reference microphone.
Let $\vd(\theta)$ denote the steering vector of the array to direction $\theta$ relative to the first (reference) microphone, which is given by 
\begin{equation}
    \vd(\theta) = [1,e^{j\phi_2(\theta)},...,e^{j\phi_M(\theta)}]^\top,
\end{equation}
where $\phi_m(\theta)$ is the phase of the received signal at the $m$th microphone with respect to the first microphone. For example, for a uniform linear array and the typical microphone indexing, we have $\phi_m(\theta) = 2\pi\cdot m\frac{\delta}{\lambda}\sin\theta$, where $\lambda$ is the wavelength of the received signal, and $\delta$ is the distance between the microphones.

{The SRP of the DS beamformer is given by
\begin{equation}
    P_{\text{DS}}(\theta;\mathbf{\Gamma}) = \vd^H(\theta) \mathbf{\Gamma} \vd(\theta),
\end{equation}
where 
\begin{equation}
    \mathbf{\Gamma} = \mathbb{E}[\vz(l)\vz^H(l) ]
\end{equation}
is the population covariance matrix. 
Since the desired source is constantly active and assumed to be at a fixed location during the entire interval, we estimate the population correlation matrix $ \mathbf{\Gamma}$ using the sample correlation matrix $\widehat{\mathbf{\Gamma}}$ by averaging }
$\vz(l)\vz^H(l)$ over multiple STFT windows. 
We divide the STFT of the signal into $L_{\text{s}}$ disjoint segments, each consisting of $L_{\text{w}}$ consecutive STFT windows, i.e., $L_{\text{STFT}} = L_{\text{s}}\cdot L_{\text{w}}$ (for more details regarding the partitioning see Section \ref{subsec: The Segments and the Interference Sources Activity}). Then, the sample correlation matrix over each segment is computed as follows
\begin{equation}
    \label{eq: general correlation matrices}
    \begin{split}
        \sGamma_i =
        \frac{1}{L_{\text{w}}}\sum_{l=(i-1)\cdot L_{\text{w}}+1}^{i\cdot L_{\text{w}}}\vz(l)\vz^H(l),  \\
    \end{split}
\end{equation}
where $i$ is the segment index. 
To obtain a full rank correlation matrix, the number of STFT windows is set to be larger than the dimension of the matrix (the number of microphones), i.e., $L_{\text{w}}\ge M$.

The incorporation of the Riemannian geometry is realized by viewing each matrix $\sGamma_i$ as a point on the HPD manifold \cite{hiai2009riemannian} 
and considering their Riemannian mean, denoted as $\sGamma_{\text{R}}$ and given by: 
\begin{equation}
    \label{eq: Riem mean as argmin}
    \sGamma_{\text{R}} = 
    \argmin_{\mathbf{\Gamma} \in \mathcal{M}} \sum_{i=1}^{L_{\text{s}}} d_{\text{R}}^2(\mathbf{\Gamma},\sGamma_i).
\end{equation}
In general, there is no closed-form solution to (\ref{eq: Riem mean as argmin}) on the HPD manifold for more than two points \cite{moakher2005differential}. 
Therefore, Algorithm \ref{Alg: Riemannian mean interations} proposed in \cite{barachant2013classification} is used to compute the Riemannian mean of the $L_{\text{s}}$ correlation matrices. 

Once $\sGamma_{\text{R}}$ is at hand, the {SRP of the} DS {beamformer}, %
 $P_\text{DS}(\theta;\sGamma_{\text{R}})$, is computed by
 \begin{equation}
 \label{eq: ML Spectrum}
     P_\text{DS}(\theta;\sGamma_{\text{R}}) = 
     \vd^H(\theta)\sGamma_{\text{R}}\vd(\theta).
 \end{equation}
 
In the case of a single desired source and assuming the direct path is dominant in the AIR, the direction to it is set as the direction achieving the maximum value of the {SRP} of the DS beamformer, i.e.,
\begin{equation}
\label{eq: theta hat from P_ML}
    \hat{\theta}=\argmax_{\theta} P_\text{DS}(\theta;\sGamma_{\text{R}}).
\end{equation}
{We note that (\ref{eq: theta hat from P_ML}) is used since the DoA estimation is based on a single STFT frequency bin.}
In the case of $N_{\text{D}}$ desired sources, the $N_{\text{D}}$ directions are set according to the $N_{\text{D}}$ strongest lobes in the {SRP}.
The algorithm for a single desired source is described in Algorithm \ref{Alg: K points}.

\begin{algorithm}
\caption{Direction estimation in the presence of multiple interference sources}
\label{Alg: K points}
\textbf{Input:} The received signal in the STFT domain $\{\vz(l)\}_{l=1}^{L_{\text{STFT}}}$ \\
\textbf{Output:} The estimated direction of the desired source $\hat{\theta}$\\
\begin{algorithmic}[1]
\STATE Divide $\{\vz(l)\}_{l=1}^{L_{\text{STFT}}}$ into $L_{\text{s}}$ consecutive segments
\STATE For each segment $i$, compute the {sample} correlation matrix $\sGamma_i$ using (\ref{eq: general correlation matrices})
\STATE Compute $\mathbf{\Gamma}_{\text{R}}$ of the set $\{\sGamma_i\}_{i=1}^{L_{\text{S}}}$ using Algorithm \ref{Alg: Riemannian mean interations}
\STATE Compute $P_\text{DS}(\theta;\mathbf{\Gamma}_{\text{R}})$ according to (\ref{eq: ML Spectrum})
\STATE Return $\hat{\theta}=\argmax_{\theta} P_\text{DS}(\theta;\mathbf{\Gamma}_{\text{R}})$
\end{algorithmic}
\end{algorithm}

As a baseline, we consider the common practice of the DS {SRP} computation, which is typically based on the sample correlation matrix over the entire interval, i.e., $P_\text{DS}(\theta;\sGamma_{\text{E}})$, where  
\begin{equation}
\label{eq: correlation matrix estimation from entire signal}
    \sGamma_{\text{E}} =
    \frac{1}{L_{\text{STFT}}}\sum_{l=1}^{L_{\text{STFT}}} \vz(l)\vz^H(l).
\end{equation}

We observe that computing the Euclidean mean of the {sample} correlation matrices per segment, $\{\sGamma_i\}_{i=1}^{L_{\text{s}}}$, results in $\sGamma_{\text{E}}$ in (\ref{eq: correlation matrix estimation from entire signal}).  
So, $\sGamma_{\text{E}}$ in (\ref{eq: correlation matrix estimation from entire signal}) is the Euclidean counterpart of $\sGamma_{\text{R}}$, the {sample} correlation matrix resulting from the Riemannian approach.  

We will show that our Riemannian approach exploits the assumption that the desired source is constantly active and at a fixed location, whereas the interference sources are intermittent. More specifically, we will show both theoretically in Section \ref{sec: Analysis} and empirically in Section \ref{sec: simulation results} that the Riemannian mean attenuates the intermittent interferences while preserving the constantly active sources. 
In contrast, the standard Euclidean mean accumulates all the sources, and as a result, the main lobe could deviate from the direction of a desired source, and even focus on an interference source.

We show in Section \ref{sec: Analysis} and Section \ref{sec: simulation results} that our proposed approach results in a {SRP} that rejects the interference sources, allowing the beamformer to extract the DoA of the desired sources.

We remark that the proposed approach only requires that the desired sources are the only sources active during the entire interval.
{The rank of the signal matrix is not known nor needs to be estimated. The number of interference sources is unknown as well. This is by virtue of the Riemannian mean.}
Unlike other works (e.g. \cite{markovich2009multichannel}), we do not need to know the activation times of each interference, nor the number of interference sources. 
Furthermore, we do not assume that there exists a segment, at which a desired source is the only active source, namely, it could always be accompanied by interference sources.

In terms of complexity, the Riemannian approach requires the computation of the Riemannian mean of the correlation matrices, which is more complex than the Euclidean mean. 
{However, the excess complexity depends on the number of microphones, which is typically not high compared to the complexity of computing the correlation matrix which depends on the number of signal samples. Consequently, the excess complexity is negligible. 
In particular, following the iterative computation of the Riemannian mean in Algorithm \ref{Alg: K points online} in Appendix \ref{subsec: Extension to a Streaming Setting}, at each iteration, the computation involves the eigenvalue decomposition of two complex matrices of dimension $M\times M$, which is $O(M^3)$, and the computation of $6$ matrix products with the complexity of $O(M^2)$. So, at each iteration, the overall excess complexity is $O(M^3)$.  We recall that for $N$ samples, the complexity of computing the sample correlation matrix is $O(M^2N)$ which is much larger than $O(M^3)$ since typically $N>>M$.  
Empirically, we found that the Algorithm \ref{Alg: K points online} converges very quickly after several iterations.
To summarize, the proposed approach leads to improved performance with only negligible additional computational cost.}
In Appendix \ref{subsec: Extension to a Streaming Setting}, we present the estimator along with an implementation for the streaming data setting.

To evaluate the performance of the proposed approach, we define the output Signal to Interference Ratio (SIR) as follows:
\begin{equation}
\label{eq: SIR defintion}
    \begin{split}
        \text{SIR}_j(\sGamma) = \frac{P(\theta^d;\sGamma)}{P(\theta^i_j;\sGamma)},
    \end{split}
\end{equation}
where $P(\theta;\sGamma)$ is the {SRP} computed using the {sample} correlation matrix $\sGamma$, $\theta^d$ is the direction of a desired source, and $\theta^i_j$ is the direction of the $j$th interference.
When using the DS beamformer, the output SIR becomes
\begin{equation}
\label{eq: SIR_j definition}
    \begin{split}
        \text{SIR}_j(\sGamma) = \frac{\vd^H(\theta^d)\sGamma\vd(\theta^d)}{\vd^H(\theta^i_j)\sGamma\vd(\theta^i_j)}.
    \end{split}
\end{equation}

This measure of performance is used because the main challenge in this setting is the presence of interference sources rather than the microphone noise.

\section{Analysis}
\label{sec: Analysis}

In this section, we analyze the proposed approach which is based on Riemannian geometry and compare it to its Euclidean counterpart. The proofs of the statements appear in the Supplementary Material (SM). %
{In the analysis, we consider the population correlation matrix of the received signal, neglecting the estimation errors stemming from the finite sample in a segment. We note that sections \ref{Proposed Approach}, \ref{sec: Extension to other Beamformers}, \ref{sec: simulation results} consider the sample correlation matrices, and only Section \ref{sec: Analysis} considers the population correlation matrix.}

We begin with a short derivation, demonstrating that Riemannian geometry preserves better the desired source subspace in comparison to Euclidean geometry. 
Consider a single desired source and assume its ATF, denoted by $\vh_0$, is a common eigenvector of all the correlation matrices per segment associated with the same eigenvalue (this assumption is made formal in Assumption \ref{assump: uncorrelated ATFs of target and interferences} in the sequel).
In this case, according to Lemma \ref{lemma: common eigenvalue equality} (see Appendix \ref{Appendix: On the Particular Choice of the Riemannian Metric}), $\vh_0$ is an eigenvector of both means and it is associated with the same eigenvalue, namely $\lambda_0(\mathbf{\Gamma}_{\text{R}}) = \lambda_0(\mathbf{\Gamma}_{\text{E}})$, where $\lambda_i(\mathbf{\Gamma})$ is the $i$th eigenvalue of $\mathbf{\Gamma}$.
In addition, all other eigenvectors span the interference and noise subspaces.
{Under the aforementioned assumption, and by the following known property of the Riemannian mean \cite{lim2012matrix}}
\begin{equation}
\label{eq: Gamma_R predeq Gamma_E}
    \mathbf{\Gamma}_{\text{R}} \preceq
    \mathbf{\Gamma}_{\text{E}}.
\end{equation}
we get that
\begin{equation}    
\label{eq: subspace R > subspace E}
\frac{\lambda_0(\mathbf{\Gamma}_{\text{R}})}{\sum_{i=1}^{M-1}\lambda_i(\mathbf{\Gamma}_{\text{R}})} \ge
    \frac{\lambda_0(\mathbf{\Gamma}_{\text{E}})}{\sum_{i=1}^{M-1}\lambda_i(\mathbf{\Gamma}_{\text{E}})},
\end{equation}
due to the equality of the numerators and (\ref{eq: Gamma_R predeq Gamma_E}).
The inequality in (\ref{eq: subspace R > subspace E}) implies that the desired source subspace is more dominant relative to the subspace of the interference and noise in the Riemannian mean compared to the Euclidean mean.

In the remainder of this section, we extend this analysis and present additional results.

\subsection{Assumptions}
\label{subsec: Analysis_Assumptions}
To make the analysis tractable, we consider a single desired source and multiple interference sources. Therefore, we simplify the notations by omitting the superscripts $(\cdot)^i$ and $(\cdot)^d$ associated with the interference sources and the desired sources and setting the index of the desired source to $0$. 

For the purpose of analysis, we make the following assumptions:

\begin{assumption}
\label{assump: uncorrelated ATFs of target and interferences}
$ \vh_0^H\vh_j = 0, \;\;\; \forall j=1,...,N_{\text{I}}$.
\end{assumption}
\begin{assumption}
\label{assump: uncorrelated ATFs between interferences}
$ \vh_l^H\vh_j = 0, \;\;\; \forall l\neq j$.
\end{assumption}

It follows from Assumption \ref{assump: uncorrelated ATFs of target and interferences} and Assumption \ref{assump: uncorrelated ATFs between interferences} that the ATFs, associated with the desired source and the interference sources are all uncorrelated. 
These are common assumptions, e.g., see  \cite{markovich2009multichannel}. 
{The assumptions are made only for analysis purposes, whereas in the experimental results, we consider acoustic signals in a reverberant environment without any assumptions.}
We note that we do not assume there exists a segment at which only one of the sources is active (e.g., as in \cite{markovich2009multichannel}). 
In case an interference source is only partially active during a segment, we consider it active during the entire segment.

The population correlation matrix of the $i$th segment is given by
\begin{equation}
\label{eq: corr matrix}
    \begin{split}
        \mathbf{\Gamma}_i = 
        \sigma_0^2\vh_0\vh_0^H +\mH\Lambda_i\mH^H +  \sigma_v^2 \mI_{M\times M},
    \end{split}
\end{equation}
where $\sigma_0^2$ and $\sigma_v^2$ are the power of the desired source and the power of the noise, respectively.
The diagonal matrix $\Lambda_i$ captures the signal power of the interference sources and is given by: 
\begin{equation}
    \begin{split}
        \Lambda_i = 
        \text{diag}\left( \sigma_1^2(i)\cdot \mathcal{I}_{i\in\mathcal{L}_1}, \;...\; , \sigma_{N_{\text{I}}}^2(i)\cdot \mathcal{I}_{i\in\mathcal{L}_{N_{\text{I}}}} \right) ,
    \end{split}
\end{equation}
where $\mH=\mH^i(l,k)$ due to the omission of the indices and $\sigma_j^2(i)= \mathbb{E}[|s_j(n)|^2|n \in  i\text{th segment} ]$ is the expected signal power of the $j$th interference source at the $i$th segment, $\mathcal{L}_j$ is the set of segments at which the $j$th interference source is active, and $\mathcal{I}_{i\in\mathcal{L}_j}$ is an indicator function, attaining the value of $1$ when the $j$th interference is active during the $i$th segment and $0$ otherwise. 
We denote by $\tau_j=\frac{|\mathcal{L}_j|}{L_{\text{s}}}$ the relative number of segments during which the $j$th interference source is active. We assume the same expected power at all the segments in the interval, i.e., $\sigma_j^2(i) = \sigma_j^2 \;$ for all $j=1,\ldots,N_{\text{I}}$ and $i=1,\ldots,L_{\text{s}}$.

We continue with defining the Signal to Noise Ratio (SNR) {at the $m$th microphone} as
\begin{equation}
\label{eq: snr definition}
    \begin{split}
        \text{SNR}{_m} = \frac{\sigma_0^2{|\vh_0[m]|^2}}{\sigma_v^2},
    \end{split}
\end{equation}
{where $|\vh_0[m]|^2$ is the attenuation of the signal due to the acoustic channel between the desired source and the $m$th microphone.} 
Since we focus on a single frequency bin, (\ref{eq: snr definition}) is the narrowband SNR. 

To capture the correlation between the steering vectors and the ATFs, we define
\begin{equation}
    \begin{split}
        \rho_{rs}       =
        \frac{|\langle \vd_r , \vh_s\rangle|^2}{\|\vd_r\|^2 \cdot \|\vh_s\|^2} = 
        \frac{|\langle \vd_r , \vh_s\rangle|^2}{M\|\vh_s\|^2},
    \end{split}
\end{equation}
where $r,s = 0,1,2,...,N_{\text{I}}$, indicating the desired source or an interference source.

We conclude the preliminaries of the analysis with two additional assumptions.
\begin{assumption}
\label{assump: fixed rhos}
$\rho_{rr}$ is fixed $\forall r$, and $\rho_{rs}$ is fixed $\forall r\neq s$.
\end{assumption}
Assumption \ref{assump: fixed rhos} implies that the correlation between the ATFs and the steering vectors depends only on whether they are associated with the same source or not. 
Following Assumption \ref{assump: fixed rhos}, henceforth we denote $\kappa = \rho_{rr}$ and $\rho = \rho_{rs}$ for $r\neq s$.
\begin{assumption}
\label{assump: rho_c>rho_d}
$\kappa > \rho$.
\end{assumption}
Assumption \ref{assump: rho_c>rho_d} is typically made in the context of source localization. It implies that the correlation between a steering vector to a source and the ATF associated with that source is higher than the correlation between a steering vector to a source and the ATF associated with a different source.

\subsection{Main Results}

Our first result states that the output SIR (\ref{eq: SIR_j definition}) of the Riemannian-based DS beamformer is higher than the output SIR of the Euclidean-based DS beamformer. 

\begin{proposition}
\label{prop: SIR and comparison}
For every interference source $j$, the following holds
\begin{equation}
    \text{SIR}_j(\mathbf{\Gamma}_{\text{R}}) > \text{SIR}_j(\mathbf{\Gamma}_{\text{E}}), \;\;\; %
\end{equation}
for any number of microphones in the array.

\end{proposition}

Examining the dependency of the output SIR on the noise power $\sigma_v^2$ leads to the following result.

\begin{proposition}
\label{prop: derivatives of Riemannian SIR}
If 
\begin{equation}
 \label{eq: condition sigma_d>c_l^R}
 \begin{split}
     \sigma_0^2\|\vh_0\|^2 \ge \sigma_j^{2}\tau_j\|\vh_j\|^2, \;\; \forall j,
 \end{split}
 \end{equation}
then
  \begin{equation} 
 \frac{\partial}{\partial\sigma_v^2} \text{SIR}_j(\mathbf{\Gamma}_{\text{R}})   < 
 \frac{\partial}{\partial\sigma_v^2} \text{SIR}_j(\mathbf{\Gamma}_{\text{E}}) < 0.
\end{equation}
\end{proposition}
Namely, the lower the noise power is the higher the SIR is, and
the improvement in $\text{SIR}_j(\mathbf{\Gamma}_{\text{R}})$ is greater than the improvement in $\text{SIR}_j(\mathbf{\Gamma}_{\text{E}})$.
Since we established that the Riemannian approach is better than the Euclidean one in terms of the SIR in Proposition \ref{prop: SIR and comparison}, Proposition \ref{prop: derivatives of Riemannian SIR} implies that increasing the SNR further increases the gap between the two approaches. Nevertheless, it also indicates that the performance of the Riemannian approach in terms of the SIR is more sensitive to noise compared to the Euclidean counterpart.
Note that this statement holds under condition (\ref{eq: condition sigma_d>c_l^R}), which implies that the received power of the desired source is stronger than the received power of each interference source, considering the attenuation stemming from the activity duration.
See more details in Section I-C in the SM. %

The proofs of Proposition \ref{prop: SIR and comparison} and Proposition \ref{prop: derivatives of Riemannian SIR} rely on the following lemma, which is important in its own right.
\begin{lemma}
\label{lemma: Riemannian Euclidean mean}
The Riemannian or the Euclidean mean of the population correlation matrices of the segments (\ref{eq: corr matrix}) over the entire interval can be written in the same parametric form as:
\begin{equation}
\begin{split}
\label{eq: Gamma expression}
    \mathbf{\Gamma} = 
     \sigma_0^2\vh_0\vh_0^H + 
    \sum_{j=1}^{N_{\text{I}}} \mu^2_j \vh_j\vh_j^H +
    \sigma_v^2\mI.
\end{split}
\end{equation}
The Riemannian mean $\mathbf{\Gamma}_{\text{R}}$ is obtained by setting the parameters $\mu_j$ to 
\begin{equation}
\label{eq: mu_j Riem}
\mu_j^2 = 
\frac{(\sigma_j^2\Vert \vh_j\Vert^2 + \sigma_v^2)^{\tau_j}(\sigma_v^2)^{1-\tau_j}-\sigma_v^2}{\Vert \vh_j\Vert^2},
\end{equation}
and the Euclidean mean $\mathbf{\Gamma}_{\text{E}}$ is obtained by setting
\begin{equation}
\label{eq: mu_j Euc}
\mu_j^2 =
 \sigma_j^2 \tau_j.
\end{equation}
\end{lemma}
We note that only assumptions \ref{assump: uncorrelated ATFs of target and interferences} and \ref{assump: uncorrelated ATFs between interferences} are necessary for this lemma to hold.
In addition, we note that if the interference sources are always active, i.e., $|\mathcal{L}_j|=L_{\text{s}},\; \forall j$, it holds that $\mathbf{\Gamma}_{\text{R}}=\mathbf{\Gamma}_{\text{E}}$.

Lemma \ref{lemma: Riemannian Euclidean mean} shows that both $\mathbf{\Gamma}_{\text{R}}$ and $\mathbf{\Gamma}_{\text{E}}$, i.e., the population correlation matrix of the segments in (\ref{eq: corr matrix}), can be decomposed into three terms associated with the desired source, the interference sources, and the noise. By (\ref{eq: Gamma expression}), the desired source term and the noise term (the first and third terms) are the same in both the Riemannian and the Euclidean means. 
In contrast, the coefficients in (\ref{eq: mu_j Riem}) and (\ref{eq: mu_j Euc}) imply that the amplitude of the interference sources term (the second term) depends on the used geometry.
By further inspecting the expressions of $\{\mu_j^2\}$, we see that the interference attenuation using the Riemannian geometry in (\ref{eq: mu_j Riem}) is more involved than its Euclidean counterpart in (\ref{eq: mu_j Euc}), depending not only on the interference power and the duration of activity but also on the noise power and the corresponding ATF.

Furthermore, considering $\mu_j$ in (\ref{eq: mu_j Euc}), the condition (\ref{eq: condition sigma_d>c_l^R}) could be viewed as the dominance of the desired source after the attenuation of the Euclidean mean. Discussion about the condition (\ref{eq: condition sigma_d>c_l^R}) for $\mu_j$ in the Riemannian case in (\ref{eq: mu_j Riem}) appears in Section I-C in the SM. %
Next, we examine a family of correlation matrices that pertain to the same parametric form as in (\ref{eq: Gamma expression}) in Lemma \ref{lemma: Riemannian Euclidean mean}, i.e.,
\begin{equation}
\label{eq: Gamma_a}
    \mathbf{\Gamma}_\va =
     \vh_0\vh_0^H + 
    \sum_{j=1}^{N_{\text{I}}} a_j \vh_j\vh_j^H +
    \sigma_v^2\mI,
\end{equation}
for some coefficients $\va = [a_1,a_2,...,a_{N_{\text{I}}}]$. Without loss of generality, we set the coefficient of $\vh_0\vh_0^H$ to $1$.
We note that $\mathbf{\Gamma}_\va$ is in accordance with (\ref{eq: corr matrix}).
For any $j$, we have that
\begin{equation}
\label{eq: Gamma_opt}
    \mathbf{\Gamma}_{\text{opt}} \equiv \argmax_{\mathbf{\Gamma}_\va}
    \text{SIR}_j(\mathbf{\Gamma}_{\va}) =
    \vh_0\vh_0^H + \sigma_v^2\mI, 
\end{equation}
where $\va = \mathbf{0}$.
Consequently, 
\begin{equation}
\label{eq: SIR(Gamma_opt)}
    \begin{split}
    \text{SIR}_j(\mathbf{\Gamma}_{\text{opt}}) = 
     \frac{\vd_0^H(\vh_0\vh_0^H + \sigma_v^2\mI)\vd_0}{\vd_j^H(\vh_0\vh_0^H + \sigma_v^2\mI)\vd_j}.
     \end{split}
\end{equation}

Considering vanishing noise, i.e., when the noise power approaches zero, the following result stems from Lemma \ref{lemma: Riemannian Euclidean mean} by considering the limit $\lim_{\sigma_v^2\rightarrow 0} \mu_j^2 = 0$ using (\ref{eq: mu_j Riem}).
\begin{corollary}
\label{cor: SIR for Gamma_R for a negligble noise}
\begin{equation}
\label{eq: Gamma_R-->Gamma_opt in limit of negligble noise}
    \begin{split}
   \lim_{\sigma_v^2\rightarrow 0} \mathbf{\Gamma}_{\text{R}} = 
   \mathbf{\Gamma}_{\text{opt}}.
    \end{split}
\end{equation}

\end{corollary}
According to Corollary \ref{cor: SIR for Gamma_R for a negligble noise}, the Riemannian mean approaches the optimal correlation matrix as the noise becomes negligible.
By adding a condition on the presence of the interference sources, from Lemma \ref{lemma: Riemannian Euclidean mean} and (\ref{eq: mu_j Riem}) we also have the following.
\begin{corollary}
\label{cor: SIR for infinite interferece power}
For any interference source $j$, if $\tau_j<\frac{1}{2}$, then
\begin{equation}
    \begin{split}
    \lim_{\sigma_j^2 \rightarrow \infty,\sigma_v^2\rightarrow 0} \mathbf{\Gamma}_{\text{R}} = 
    \mathbf{\Gamma}_{\text{opt}}.
    \end{split}
\end{equation}
Additionally, if $\tau_j<\frac{1}{2}$ for all $j$, then 
\begin{equation}
    \begin{split}
    \lim_{\sigma_j^2 \rightarrow \infty \forall j=1,\ldots, N_{\text{I}},\sigma_v^2\rightarrow 0} \mathbf{\Gamma}_{\text{R}} = 
    \mathbf{\Gamma}_{\text{opt}}.
    \end{split}
\end{equation}

\end{corollary}

Corollary \ref{cor: SIR for infinite interferece power} implies that for vanishing noise, even when all the interference sources have infinite power, the desired source is still the dominant source in the {SRP} of the DS beamformer using the Riemannian mean. 
Following (\ref{eq: SIR(Gamma_opt)}) it holds that $\lim_{\sigma_j^2 \rightarrow \infty,\sigma_v^2\rightarrow 0}\text{SIR}_j(\mathbf{\Gamma}_{\text{opt}}) = \frac{\kappa}{\rho} > 1$ for all $j$.
For the Euclidean mean it holds that $\lim_{\sigma_j^2 \rightarrow \infty,\sigma_v^2\rightarrow 0}\text{SIR}_j(\mathbf{\Gamma}_{\text{E}}) = \frac{\rho}{\kappa} < 1 <
\frac{\kappa}{\rho} =
\lim_{\sigma_j^2 \rightarrow \infty,\sigma_v^2\rightarrow 0}\text{SIR}_j(\mathbf{\Gamma}_{\text{opt}}) $ for all $j$.
Note that we consider noise power approaching zero rather than strictly zero, because when $\sigma_v^2=0$, the correlation matrix is singular, and therefore, lies outside the HPD manifold. Additionally, in practice, noise is always present.

To illustrate the obtained expressions for the Riemannian and the Euclidean SIR, we present the following simple example.
\begin{example}
Consider an anechoic environment without attenuation, for which $\kappa = 1$ and $\rho = 0$, and two interference sources. Each interference source is active at a different segment, i.e., $L_{\text{s}}=2$, $\mathcal{L}_1 = \{1\}$, and $\mathcal{L}_2 = \{2\}$. All the sources have the same power. 
In this setting, for the Riemannian geometry, we have
\begin{equation}
    \text{SIR}_j(\mathbf{\Gamma}_{\text{R}}) = \sqrt{\frac{M}{\sigma_v^2}+1},
\end{equation}
and for Euclidean geometry, we have:
\begin{equation}
    \text{SIR}(\mathbf{\Gamma}_{\text{E}}) = 
        \frac{
        2(M + \sigma_v^2 )
        }
        {
        M + 2\sigma_v^2
        }.
\end{equation}
Therefore, in the limit of $\sigma_v^2\rightarrow 0$, or $M\rightarrow\infty$, we have $\text{SIR}(\mathbf{\Gamma}_{\text{R}}) = \infty$, whereas $\text{SIR}(\mathbf{\Gamma}_{\text{E}}) \approx 2$.
\end{example}

We conclude this analysis with a few remarks.
First, we note that $\mathbf{\Gamma}_{\text{E}}$ leads to the ML estimator by taking $\hat{\theta}_0=\argmax_{\theta} P_{\text{DS}}(\theta;\mathbf{\Gamma}_{\text{E}})$ for the interference-free setting. In this case, the Riemannian mean coincides with the Euclidean mean, i.e., the proposed method coincides with the ML estimator. The main advantage of the proposed method lies in attenuating the interference sources while preserving the desired source.
Second, under assumptions \ref{assump: uncorrelated ATFs of target and interferences} and \ref{assump: uncorrelated ATFs between interferences}, the number of sources, both desired and interference, are limited by the number of microphones in the array, i.e., $N_{\text{I}} + N_{\text{D}} < M$. In Section \ref{sec: Analysis For Correlated Multiple Interferences} we alleviate Assumption \ref{assump: uncorrelated ATFs between interferences}, which removes this restriction.
Third, following the same techniques in the proof of Proposition \ref{prop: SIR and comparison} and Proposition \ref{prop: derivatives of Riemannian SIR}, similar results are derived for an alternative definition of the SIR: $\text{SIR}_\text{tot}(\mathbf{\Gamma}) \equiv \frac{\vd_0^H\mathbf{\Gamma}\vd_0}{\sum_{j=1}^{N_{\text{I}}}\vd_j^H\mathbf{\Gamma} \vd_j}$, which captures the ratio between the desired source and the \emph{sum} of all interference sources. See Section II in the SM %
for more details.

\subsection{Relation to Signal Enhancement}

\label{sec: Analysis For Correlated Multiple Interferences}
For signal enhancement in reverberant environments, the estimation of the ATF of the desired source is typically required. In our setting, there is no segment at which the desired source is the only active source, and therefore, the ATF estimation is done in the presence of the interference sources. In such a case, the following quantity could be of interest: 
\begin{equation}
\label{eq: SIR definition with h}
    \begin{split}
         \overline{\text{SIR}}_j(\mathbf{\Gamma})= 
          \frac{\vh_0^H\mathbf{\Gamma}\vh_0}{\vh_j^H\mathbf{\Gamma}\vh_j},
    \end{split}
\end{equation}
which is different than (\ref{eq: SIR_j definition}) in the use of the ATFs instead of the steering vectors.

Similarly to Proposition \ref{prop: SIR and comparison}, the following Proposition \ref{prop: general SIR  Gamma_R superiority} examines the performance in terms of the SIR defined in (\ref{eq: SIR definition with h}). Here, assumptions \ref{assump: uncorrelated ATFs between interferences}-\ref{assump: rho_c>rho_d} are not required, and therefore, the ATFs of the interference sources could be correlated, and the number of sources is not limited by the number of microphones in the array. 
\begin{proposition}
\label{prop: general SIR  Gamma_R superiority}
Under Assumption 1, for all $j$ we have:
\begin{equation}
    \begin{split}
        \overline{\text{SIR}}_j(\mathbf{\Gamma}_{\text{R}}) \ge
        \overline{\text{SIR}}_j(\mathbf{\Gamma}_{\text{E}}). 
    \end{split}
\end{equation}
\end{proposition}

Another interesting component in signal enhancement is the Relative Transfer Function (RTF) between different microphones \cite{cohen2004relative,talmon2009relative,gannot2001signal}. We compute the RTFs with respect to the first microphone, i.e., $\frac{\vh_j}{\vh_j(1)}$.
Since the RTFs are proportional to the ATFs, assumption \ref{assump: uncorrelated ATFs of target and interferences} holds for the RTFs, and therefore, all the derived results apply to the RTFs as well.

\subsection{The Segments and the Interference Sources Activity}
\label{subsec: The Segments and the Interference Sources Activity}

In this section, we investigate the effect of misalignment between the segments and the activity of the interference sources. We consider two interference sources and two segments. We denote by $\alpha$ the offset between the segments and the activity of the interference sources. For simplicity, we consider alternately active interference sources. Suppose the first interference source is active during $\alpha\in [0,1]$ of the first segment and during $1-\alpha$ of the second segment, and suppose the second interference source is active during $1-\alpha$ and $\alpha$ of the first and second segments, respectively. %
In this case, the correlation matrices of the two segments are given by:
\begin{equation}
\label{eq: Gamma_1 and Gamma_2 with offset alpha}
\begin{split}
    \mathbf{\Gamma}_1(\alpha) &=
    \sigma_0^2\vh_0\vh_0^H + \alpha^2\sigma_1^2\vh_1\vh_1^H +
    (1-\alpha)^2\sigma_2^2\vh_2\vh_2^H +
    \sigma_v^2 \mI \\
    \mathbf{\Gamma}_2(\alpha) &=
    \sigma_0^2\vh_0\vh_0^H + 
    (1-\alpha)^2\sigma_1^2\vh_1\vh_1^H +
    \alpha^2\sigma_2^2\vh_2\vh_2^H +
    \sigma_v^2 \mI.
\end{split}
\end{equation}
The correlation matrices in (\ref{eq: Gamma_1 and Gamma_2 with offset alpha}) depend on $\alpha$, and as a result, their Riemannian mean $\mathbf{\Gamma}_{\text{R}}(\alpha)$ and their Euclidean mean $\mathbf{\Gamma}_{\text{E}}(\alpha)$ depend on $\alpha$ as well. 

Examining the dependency of the SIR on $\alpha$ leads to the following result.

\begin{proposition}
\label{prop: optimal offset for partition}
For any $\alpha\in[0,1]$, we have
\begin{equation}
\label{eq: SIR_R(alpha) > SIR_E(alpha)}
    \text{SIR}(\mathbf{\Gamma}_{\text{R}}(\alpha)) \ge \text{SIR}(\mathbf{\Gamma}_{\text{E}}(\alpha)). %
\end{equation}

\end{proposition}
Proposition \ref{prop: optimal offset for partition} states that for every misalignment between the segments and the activity of the interference sources, the Riemannian mean leads to higher SIR in comparison to its Euclidean counterpart. Equality in (\ref{eq: SIR_R(alpha) > SIR_E(alpha)}) is obtained for $\alpha=\frac{1}{2}$, which means $50\%$ offset. In this case, it holds that $\mathbf{\Gamma}_1 = \mathbf{\Gamma}_2$, and both means are the same.

Empirically, we found that the advantage of the Riemannian mean over the Euclidean mean decreases as the offset between the segments and the activity of the interference sources increases.
We leave the question of optimal partitioning of the STFT windows into segments to future work.

\section{Extension to other {DoA estimation methods}}

\label{sec: Extension to other Beamformers}
In this section, to broaden its applicability, we demonstrate the incorporation of the Riemannian approach in other beamformers.
Each beamformer generates a {spatial spectrum} from which the directions to the desired sources are estimated according to the highest peaks in the {spectrum}. 

As a subspace (SbSp) approach, we implement MUSIC \cite{schmidt1986multiple} in the following way. 
given $N_{\text{D}}$ desired sources, we take the leading $N_{\text{D}}$ eigenvectors of the {sample} correlation matrix, $\sGamma$, and construct the signal subspace matrix, $\mU(\sGamma)\in\sC^{M\times N_{\text{D}}}$, whose columns are the ${N}_{\text{D}}$ eigenvectors. Then, the SbSp {spectrum} is defined by 
\begin{equation}
\label{eq: SS spectrum}
    \mP_{\text{SbSp}}(\theta;\sGamma) = \vd^H(\theta)\mU(\sGamma)\mU^H(\sGamma)\vd(\theta)   .
\end{equation}
We note that the appropriate number of eigenvectors $N_{\text{D}}$ (the dimension of the subspace) needs to be estimated.

Similarly to the DS beamformer based on the {SRP} in (\ref{eq: ML Spectrum}) and (\ref{eq: theta hat from P_ML}), the Riemannian and the Euclidean SbSp methods are given by $\mP_{\text{SbSp}}(\theta;\sGamma_{\text{R}})$ and $\mP_{\text{SbSp}}(\theta;\sGamma_{\text{E}})$, respectively, according to (\ref{eq: SS spectrum}).
The SbSp method could also benefit from our Riemannian approach.
Recalling assumptions \ref{assump: uncorrelated ATFs of target and interferences} and \ref{assump: uncorrelated ATFs between interferences} and the structure of the mean correlation matrices $\mGamma_{\text{R}}$ and $\mGamma_{\text{E}}$ in (\ref{eq: Gamma expression}), we see that $\vh_0$ is an eigenvector of both $\mGamma_{\text{R}}$ and $\mGamma_{\text{E}}$, spanning the signal subspace (we assume a single desired source for simplicity). Moreover, the vectors $\{\vh_j\}$ are also eigenvectors of $\mGamma_{\text{R}}$ and $\mGamma_{\text{E}}$, spanning the subspace of the interference sources. SbSp methods typically focus on the principal eigenvector. Following (\ref{eq: Gamma expression}), the principal eigenvector is determined according to the largest coefficient among $\sigma_0$ and $\{\mu_j\}_{j=1}^{N_{\text{I}}}$. The parameter $\mu_j$ in (\ref{eq: mu_j Riem}) for the Riemannian mean $\mGamma_{\text{R}}$ is smaller than $\mu_j$ in (\ref{eq: mu_j Euc}) for the Euclidean mean $\mGamma_{\text{E}}$, as proven in Proposition \ref{prop: SIR and comparison}. As a result, the signal subspace is more dominant relative to the interference subspace when considering the Riemannian mean in comparison to the Euclidean mean, implying better results for the Riemannian SbSp approach.
For an interference source with sufficiently high power, the leading eigenvector of $\mathbf{\Gamma}_{\text{E}}$ could span the interference subspace rather than the signal subspace, whereas the leading eigenvector of $\mathbf{\Gamma}_{\text{R}}$ spans the signal subspace.
In Section \ref{sec: simulation results}, we demonstrate these SbSp methods empirically and show the advantage of the Riemannian approach. 

Our approach is also applicable to the MVDR beamformer \cite{capon1969high}, whose {spectrum} is given by
\begin{equation}
\label{eq: MVDR spectrum}
    \mP_{\text{MVDR}}(\theta;\sGamma) = \frac{1}{\vd^H(\theta)\sGamma^{-1}\vd(\theta)}.
\end{equation}
The typical {spectrum} of the MVDR beamformer is obtained by using $\sGamma_{\text{E}}$ in (\ref{eq: MVDR spectrum}), namely $\mP_{\text{MVDR}}(\theta;\sGamma_{\text{E}})$. 
We propose to use the Riemannian mean by setting $\sGamma$ to be $\sGamma_{\text{R}}$ in (\ref{eq: MVDR spectrum}) to obtain $\mP_{\text{MVDR}}(\theta;\sGamma_{\text{R}})$.

%

%
%

%


%

%

{In principle, many DoA estimation methods that employ the sample correlation matrix could potentially benefit from the proposed approach, even if it is not based on a beamformer. For example, the Bayesian learning method for signal recovery for DoA estimation proposed in \cite{hu2016source} employs the sample correlation matrix. Its Riemannian alternative is implemented by following their algorithm only with the Riemannian mean instead of the Euclidean mean. The empirical results appear in Section \ref{sec: simulation results}}.

\section{Simulation Results}
\label{sec: simulation results}
In this section, we demonstrate the performance of the proposed approach based on Riemannian geometry, and compare it to Euclidean geometry, implicitly considered by the common practice\footnote{The code is available at  \url{https://github.com/amitaybar/Interference-Rejection-using-Riemannian-Geometry-for-DoA-Estimation}}. Additionally, we compare our approach to a heuristic method, based on the intersection of subspaces. The intersection leads to the rejection of non-common components, such as the interference sources subspace, and preserves common components, such as the desired source subspace. We refer to it as the intersection beamformer (see Appendix \ref{Apppendix: The Intersection Method} for more details).

We consider a reverberant enclosure of dimensions $5\text{m}\times 4\text{m} \times 3.5\text{m}$ consisting of a microphone array with $\text{M} = 12$ microphones.
The AIRs between the different sources and the array are generated based on the image method \cite{allen1979image}, as implemented by the simulator in \cite{habets2008room}. The sampling frequency is $16\text{KHz}$, and the length of the AIRs is set to $2048$ samples.
The received signal is transformed to the time-frequency domain using STFT with a window size of $1024$ samples with $50\%$ overlap. We test all methods using a single frequency bin, of index $250$, chosen according to the microphone spacing, which is $4.36\text{cm}$.
Here, the correlation matrix estimation is based on $16$ STFT windows, which results in a segment duration of $1.024\text{s}$.  
The emitted signals are generated as white Gaussian noise. The desired source is constantly active, whereas the interference sources are active only intermittently.

\begin{figure}
 \begin{center}
\subfloat[]
{\includegraphics[width=0.48\linewidth]{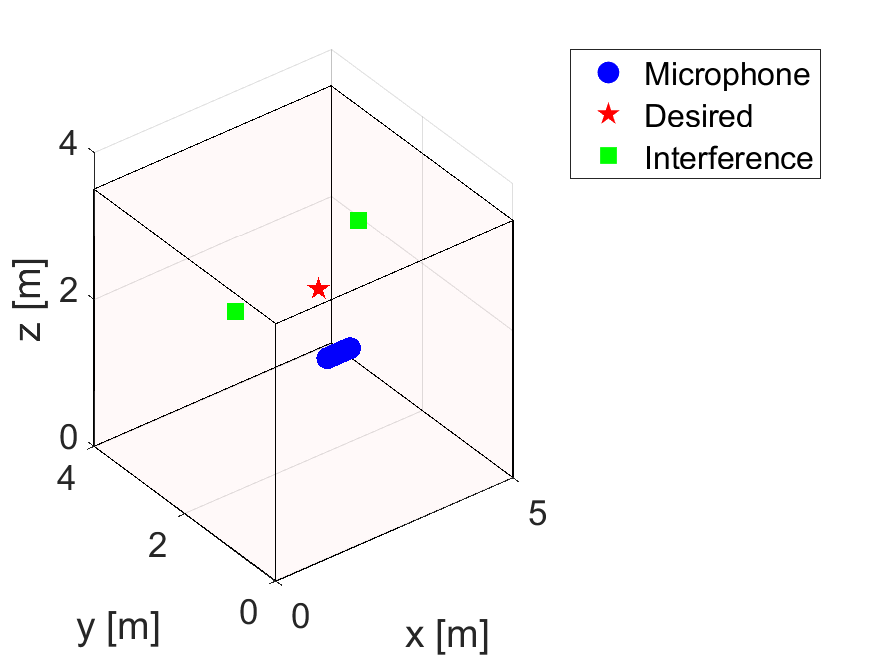}}
\subfloat[]
{\includegraphics[width=0.48\linewidth]{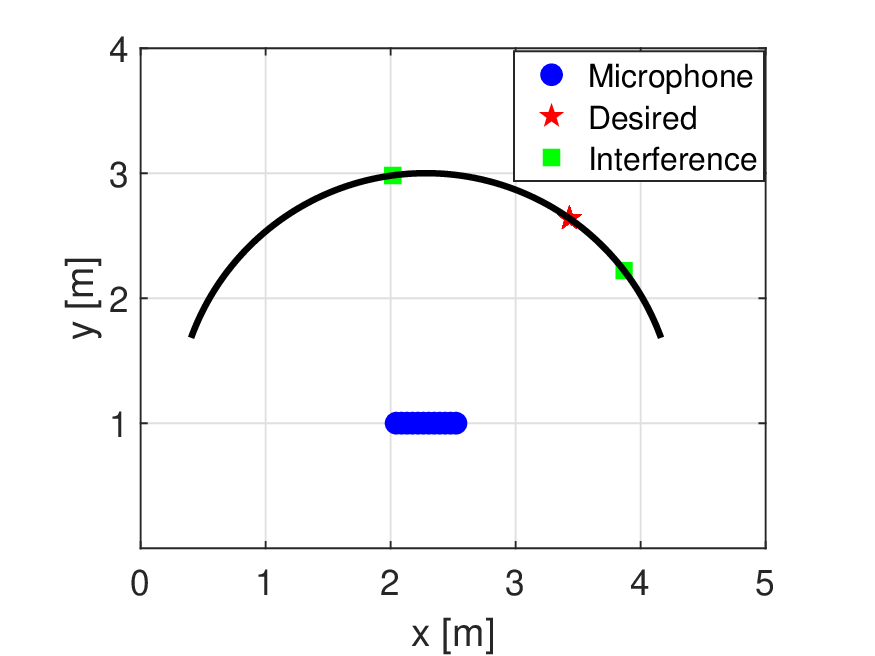}}
\caption{The reverberant room with the microphone array (blue circles), the desired source (red star), and the interference sources (green squares). (a) A 3D view. (b) A 2D view.}
\label{Fig: the scenario}
\end{center}
\end{figure}

All the sources are positioned on a $140^\circ$ arc of radius $2\text{m}$ from the center of the array on the XY plane. The heights of the interference sources vary randomly, uniformly distributed between $0.5\text{m}$ and $3\text{m}$. The height of the desired source is set to $1.8 \text{m}$.
Figure \ref{Fig: the scenario} presents the room layout, where the (two) interferences are marked by green squares, the desired source by a red star, and the microphone array by blue circles. The leftmost microphone is positioned at $(2.0436\text{m} ,1\text{m} ,2\text{m})$, and the rest are positioned $4.36\text{cm}$ apart along the x-axis.
While we focus on this specific configuration, we note that we tested other configurations that yielded similar results.

We examine the performance of both the DS and the SbSp methods. Algorithm \ref{Alg: K points} is used for the proposed Riemannian DS, and the common practice is implemented by replacing step $3$ in Algorithm \ref{Alg: K points} with (\ref{eq: correlation matrix estimation from entire signal}).
The SbSp methods require knowing the dimension of the signal space of the mean correlation matrix. 
For the intersection method only, the dimension of the signal space of each segment is also required. 
Since the desired source may not be the strongest source received at the array, we need to consider all the active sources, and not merely the strongest one when estimating the dimension of the signal subspace. To estimate the dimension, we implement a heuristic algorithm, based on the spectral gap. We consider the dimension of the signal space to be the number of eigenvalues higher than a threshold. For all methods, the threshold for the mean correlation matrix is the mean value plus the standard deviation of the eigenvalues (normalized to a unit sum). For the intersection method, the threshold for the {sample} correlation matrix of each segment is $1.5$ times the mean of the eigenvalues of the {sample} correlation matrix. Apart from this practical implementation, we also present results for an oracle implementation, assuming the dimension of the signal space is perfectly known.

For quantitative evaluation, {we use the root mean square error (RMSE) and the accuracy, for which DoA estimation error that is smaller than $3^\circ$ (arbitrarily chosen) is considered accurate. In addition,} we use two {other} metrics. The first is the mean of the empirical output SIR with respect to all the interference sources, which is given by:
\begin{equation}
    \text{SIR} = \frac{1}{N_{\text{I}}}\sum_{j=1}^{N_{\text{I}}} \frac{P({\theta}^d)}{P(\theta^i_j)}, 
\end{equation}
where $P$ is the {spectrum} computed using the evaluated method.
The second metric is the directivity \cite[ch.2]{van2004optimum}, which is given by 
\begin{equation}
    \mathcal{D}(\sGamma) = \frac{P({\theta}^d)}{\frac{1}{2}\int_0^\pi P(\theta)\sin(\theta)d\theta}.
\end{equation}

{The proposed approach results in inherent interference rejection,  which is its greatest merit. The attenuation of the interference sources by our approach allows for accurate DoA estimation even in the presence of strong interference sources. The measure of SIR is indicative of the amount of interference rejection}.

In the first experiment, we consider two interference sources, each active at a single, but disjoint, segment, resulting in a signal of $2.048\text{s}$. The reverberation time is set to $150\text{ms}$, and the SNR is ${2}0\text{dB}$.

\begin{figure}
 \begin{center}
 \subfloat[]
 {
\includegraphics[width=0.7\linewidth]{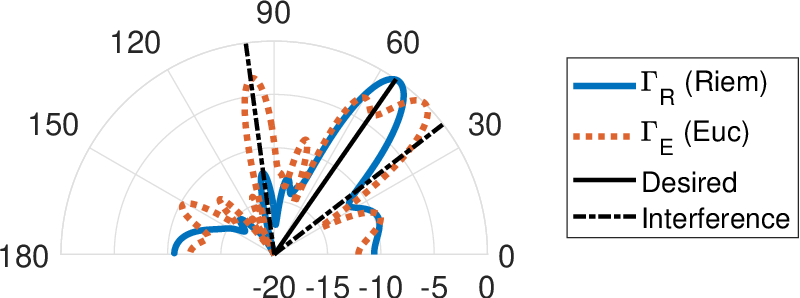}
}\\
\subfloat[]
 {
\includegraphics[width=0.7\linewidth]{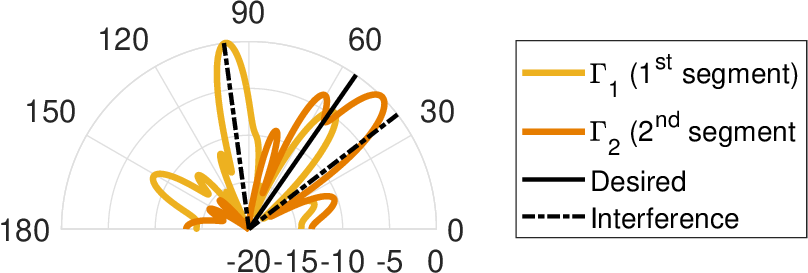}
}
\caption{The {SRP} of the DS beamformer using (a) $\sGamma_{\text{R}}$ in solid blue and $\sGamma_{\text{E}}$ in dashed red, and (b) $\sGamma_1$ and $\sGamma_2$ in different shades of orange. The black solid line indicates the direction of the desired source, and the dashed black lines indicate the directions to the interference sources. The Input SIR is $-6\text{dB}$.}
\label{Fig: Polar plot 2 interferences -6db SIR 50dB SNR}
\end{center}
\end{figure}

We start with an example of the {SRP} of the DS beamformer (see (\ref{eq: ML Spectrum})), computed using $\sGamma_{\text{R}}$ and $\sGamma_{\text{E}}$, which is presented at the top of Figure \ref{Fig: Polar plot 2 interferences -6db SIR 50dB SNR}. The {SRP} of the Riemannian DS beamformer is shown in solid blue and the {SRP} of the Euclidean DS beamformer is shown in dashed red. Both {SRP}s are in a dB (log) scale. The directions to the desired source and the interference sources are represented by a black solid line and a dashed black line, respectively. We see that by using $\sGamma_{\text{R}}$ the main lobe is directed towards the desired source. In contrast, the {SRP} using $\sGamma_{\text{E}}$ is peaked at $2$ different directions, none of which is the direction to the desired source. The bottom of Figure \ref{Fig: Polar plot 2 interferences -6db SIR 50dB SNR} is the same as the top, only with the {SRP} computed using the {sample} correlation matrix of each of the two segments, presented in different shades of orange. Even though the main lobes of the two {SRP}s are not pointing toward the desired source, the Riemannian mean leads to a {SRP} with the main lobe directed at the desired source, whereas the lobes to other directions are highly attenuated. We emphasize that viewing the {SRP} of the Euclidean DS beamformer in addition to the {SRP} of each segment separately does not allow correct identification of the desired source. 
{We note that since the population correlation matrix is approximately low rank and the number of microphones is typically small, the estimation of the correlation matrix based on a small sample is feasible.}

Next, we randomly generate $200$ different pairs of positions for the interference sources. For each pair, the desired source is located at $20$ different equally spaced directions along the arc (with the height of $1.8\text{m}$). Thus, in total, $4000$ different scenarios are examined. 

\begin{figure}
 \begin{center}
 \subfloat[]
{
\includegraphics[width=1\linewidth]{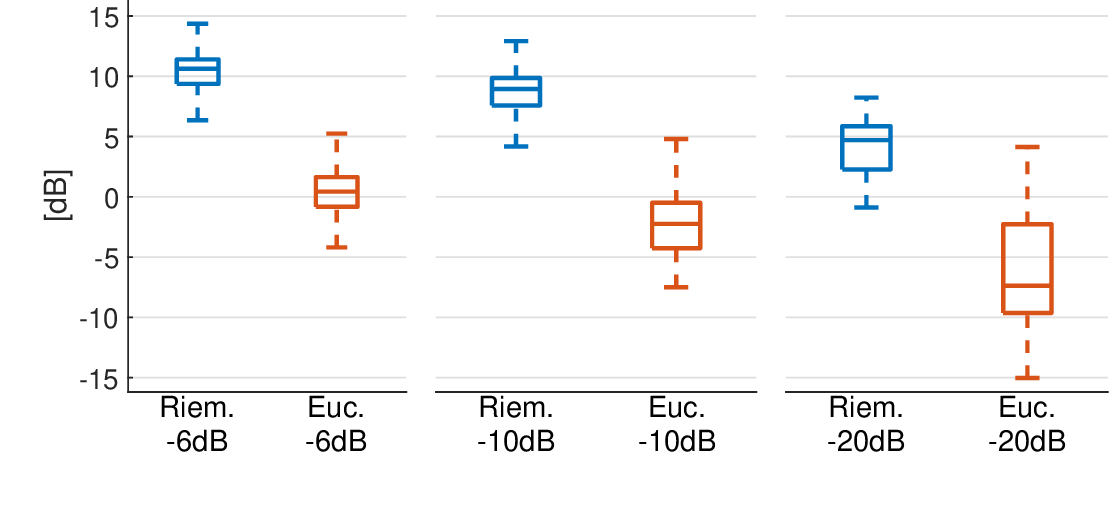}
}\\
\subfloat[]
{
\includegraphics[width=1\linewidth]{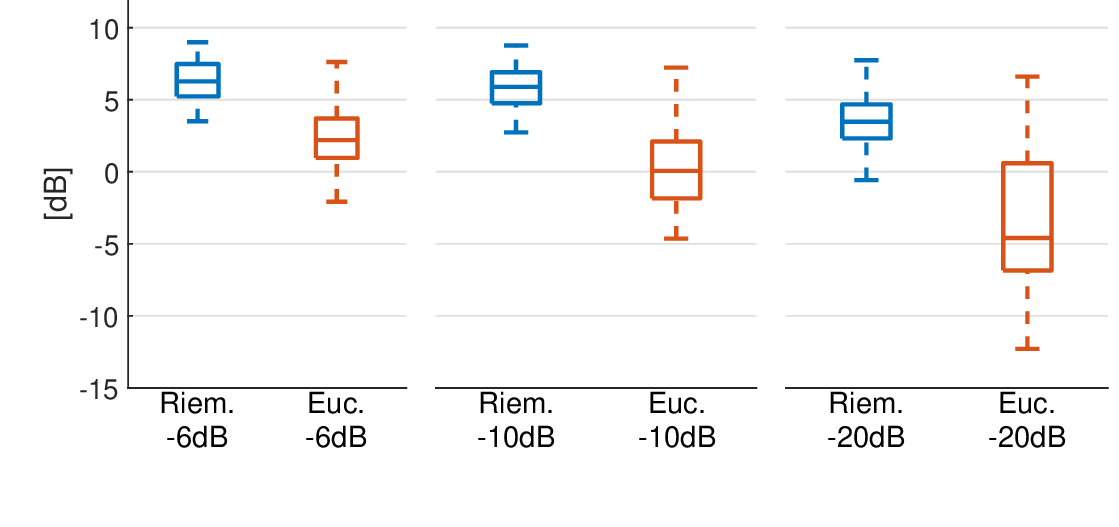}
}
\caption{(a) The mean output SIR and (b) the directivity for two interference sources, for the Riemannian and the Euclidean DS method. The x-axis indicates the input SIR, and the y-axis indicates the output SIR. The box indicates the $25$th and $75$th percentiles, and the central line marks the median. Several input SIR values are presented.}
\label{Fig: Mean SIR Directivity 2 interferences ML}
\end{center}
\end{figure}

Figure \ref{Fig: Mean SIR Directivity 2 interferences ML} presents the mean output SIR (top) and the directivity (bottom) for the DS method using the correlation matrix estimates: $\sGamma_{\text{R}}$ (based on Riemannian geometry) in blue and $\sGamma_{\text{E}}$ (based on Euclidean geometry) in red. The box indicates the $25$th and $75$th percentiles, and the line marks the median. We test the different methods, Riemannian or Euclidean, in varying input SIR values, i.e. the SIR with respect to the source's power (excluding the AIRs and the beamformer processing).

We see that the Riemannian DS method attains high output SIR values, even for strong interference sources (high input SIR). In contrast, the Euclidean DS method results in relatively low output SIR values. The gap in the output SIR values between the Riemannian DS, and the Euclidean DS is up to $10\text{dB}$. These results coincide with Proposition \ref{prop: SIR and comparison}, stating that $\text{SIR}_j(\mathbf{\Gamma}_{\text{R}}) > \text{SIR}_j(\mathbf{\Gamma}_{\text{E}})$, for every interference source $j$. 
We emphasize that the Euclidean mean, $\sGamma_{\text{E}}$, is equivalent to the common practice of using the entire signal for a single correlation matrix estimation.
Since both the mean output SIR and the directivity present similar trends, and due to space considerations, in the following, we only present the mean output SIR. 

\begin{figure}[t]
 \begin{center}
 \subfloat[]
 {
\includegraphics[width=1\linewidth]{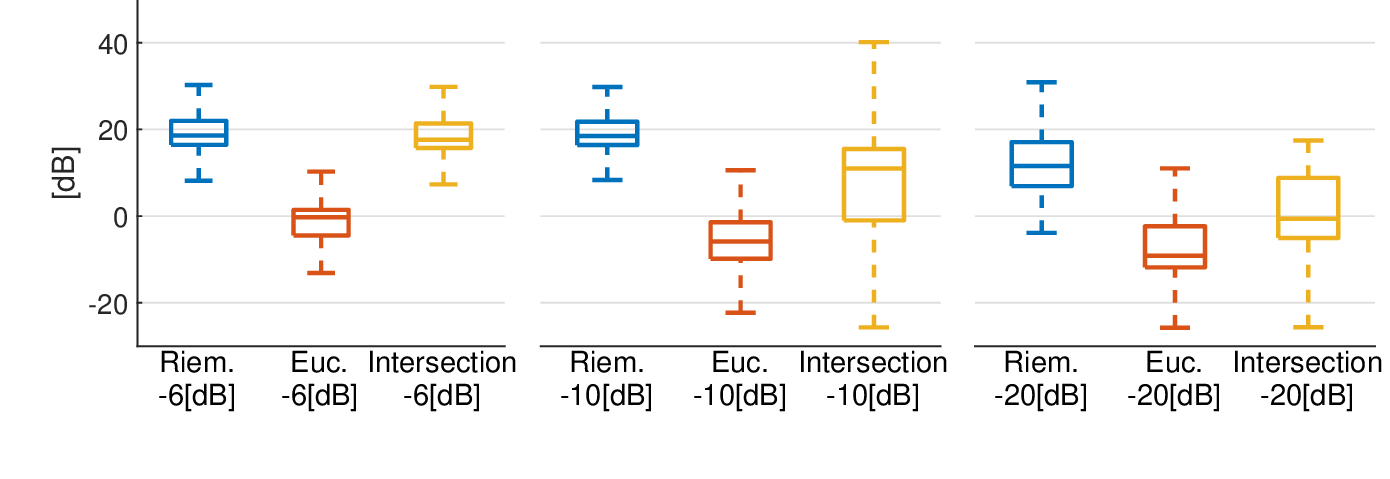}
}\\ 
\subfloat[]
{
\includegraphics[width=1\linewidth]{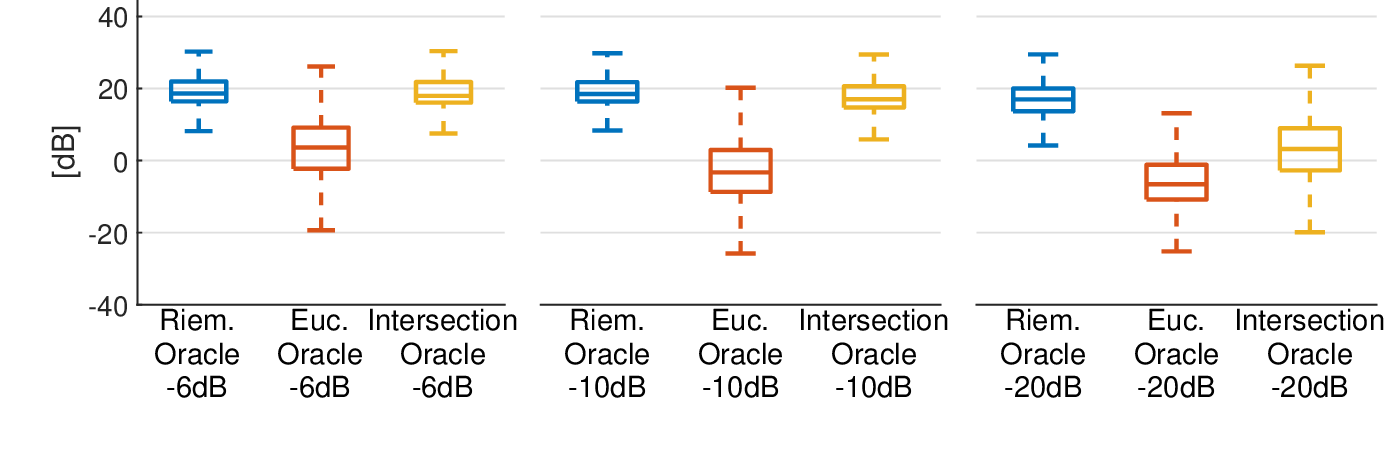}
}
\caption{The mean output SIR of the SbSp methods. (a) Practical implementation. (b) Using an oracle. The box indicates the $25$th and $75$th percentiles and the central line marks the median. The x-axis indicates the input SIR, and the y-axis indicates the output SIR. Several input SIR values are presented.}
\label{Fig: Mean SIR Directivity 2 interferences SS}
\end{center}
\end{figure}

Figure \ref{Fig: Mean SIR Directivity 2 interferences SS} is the same as Figure \ref{Fig: Mean SIR Directivity 2 interferences ML}, but presenting the SbSp method with the addition of the intersection method, which appears in orange. The top subfigure presents the results for the practical implementation that includes estimating the dimension, whereas the bottom subfigure presents the results for the oracle. %
We see that the Riemannian approach outperforms its Euclidean counterpart, by approximately $20\text{dB}$. In addition, the oracle SbSp method is better than the practical SbSp.
In comparison to Figure \ref{Fig: Mean SIR Directivity 2 interferences ML}(top), it can be seen that the Riemannian SbSp method results in higher output SIRs than the Riemannian DS method. In contrast, for the Euclidean approach, the SbSp method yields slightly lower SIRs than the DS method. The reason is that the Riemannian mean better attenuates the interference sources, allowing for a better estimation of the signal subspace than the Euclidean mean.

\begin{figure}
 \begin{center}
 \subfloat[]
 {
\includegraphics[width=0.49\linewidth]{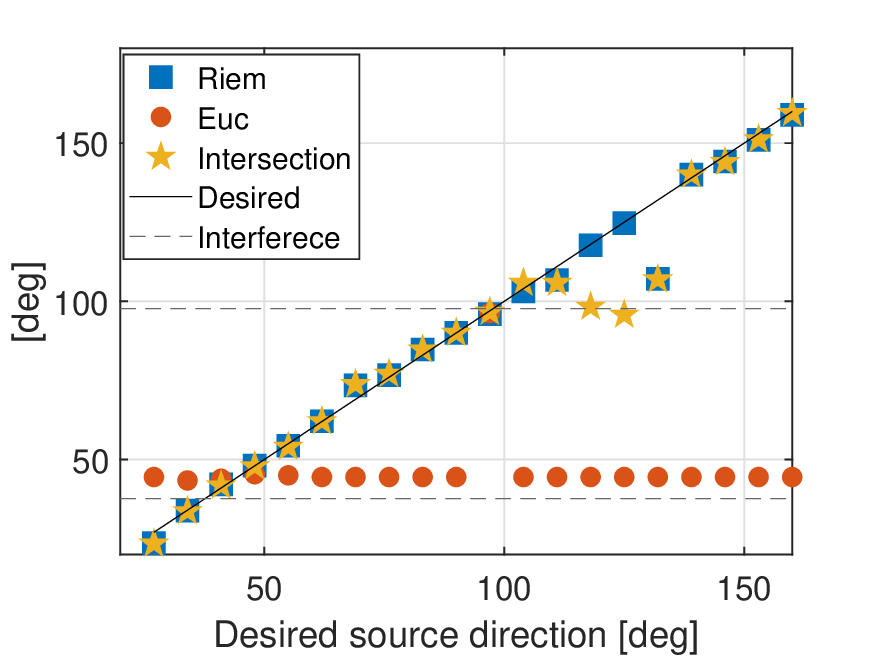}
}
\subfloat[]
{
\includegraphics[width=0.49\linewidth]{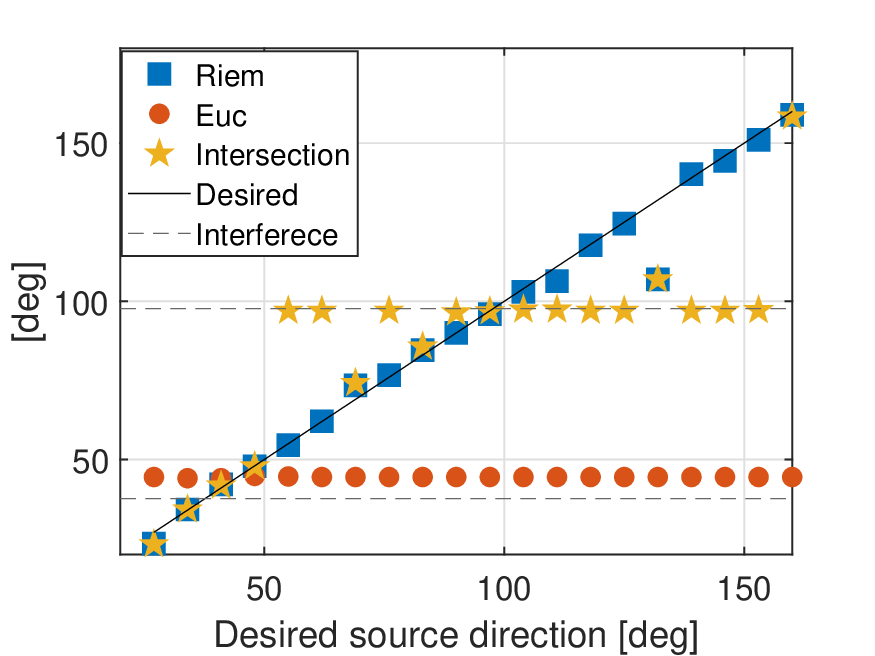}
}
\caption{Estimation of the DoA to the desired source for (a) input SIR of $-6\text{dB}$ and (b) input SIR of $-10\text{dB}$. 
}
\label{Fig: Direction estimation and error 2 interferences -6db SIR 40dB SNR}
\end{center}
\end{figure}

We continue with examining the direction estimation to the desired source. The estimated direction is defined as the direction leading to the maximal value of the {SRP}, namely
\begin{equation}
\label{eq: theta_d as argmax of P}
    \hat{\theta}^d = 
    \argmax_\theta P(\theta).
\end{equation}
Figure \ref{Fig: Direction estimation and error 2 interferences -6db SIR 40dB SNR} shows the estimated direction to the desired source for the Riemannian DS method (blue square), the Euclidean DS method (red circle), and the intersection method (orange star). The solid black line marks the true location of the target source (at $20$ different positions). The dashed line marks the fixed location of the interference sources. 
The results for input SIR $-6\text{dB}$ and $-10\text{dB}$ appear on the left and right, respectively.
We see that using the Riemannian mean, the direction estimation follows the desired source. In contrast, using the Euclidean mean (the entire signal) results in estimating the direction of one of the interference sources. %
The intersection method is also inferior to the proposed approach, resulting in direction estimation to the desired source or an interference source depending on the input SIR.
{We note that since the DoA is estimated using (\ref{eq: theta_d as argmax of P}), only the main lobe of the SRP is considered for the DoA estimation (for simplicity, we assume a single desired source). Typically-used beamformers, based on Euclidean geometry, result in a main lobe that is pointing toward the strongest source which is an interference source. Since the number of sources is unknown, and their relative positioning is unknown as well, considering even lower lobes does not allow for estimating the DoA of the desired source using the SRP. Even if the interference source does not mask the desired source (which is often the case), we cannot differentiate between the interference source and the desired source, and cannot assign the correct direction to the desired source. In contrast, the Riemannian approach rejects the interference sources, resulting in a main lobe that points toward the desired source only.}
{We report that similar results as in Figure \ref{Fig: Direction estimation and error 2 interferences -6db SIR 40dB SNR} are obtained also for an SNR value of $0\text{dB}$ and for interference sources with similar DoA. Results for speech signals from the TIMIT dataset appear in Appendix \ref{Apppendix: additional experimental results} where similar trends are demonstrated. }

\begin{figure}
 \begin{center}
\subfloat[]
{
\includegraphics[width=0.49\linewidth]{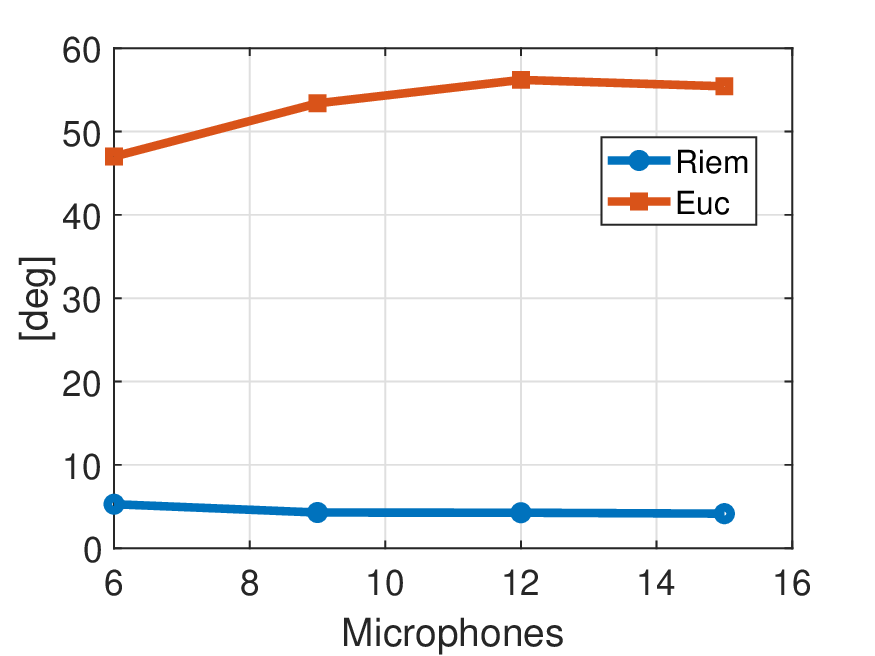}
}  
\subfloat[]
{
\includegraphics[width=0.49\linewidth]{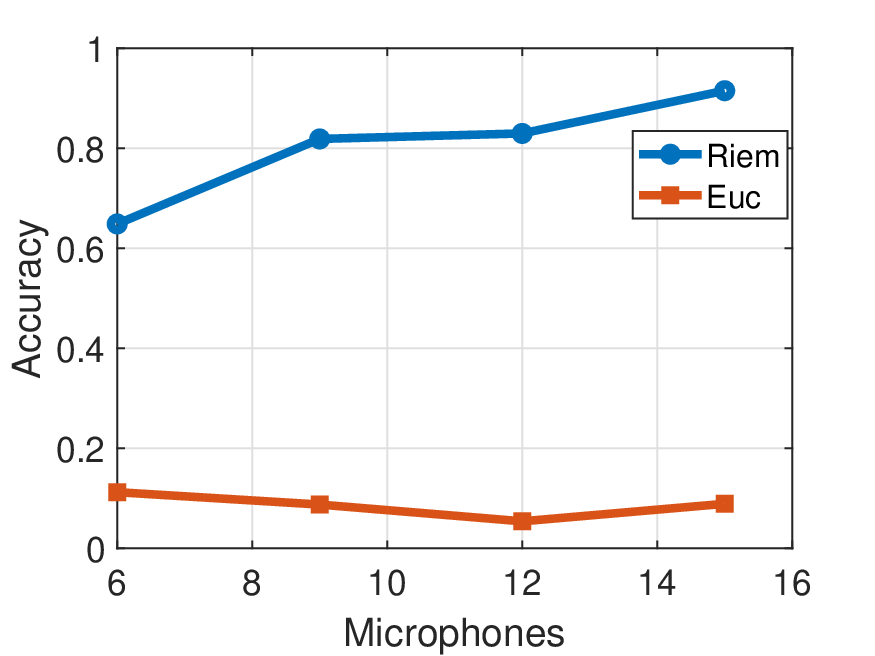}
}
\caption{{(a) RMSE and (b) accuracy versus the number of microphones for two interference sources with input SIR of $-10$dB.}}
\label{fig: RMSE Accuracy vs STFT windows and microphones}
\end{center}
\end{figure}

{We repeat the experiment for $200$ iterations, where at each iteration the desired source is positioned at $20$ different positions as before. Figure \ref{fig: RMSE Accuracy vs STFT windows and microphones} presents the RMSE and the accuracy for a different number of microphones for the Riemannian approach (blue circles) and its Euclidean counterpart (red squares).
We recall that DoA estimation error that
is smaller than $3^\circ$ (arbitrarily chosen) is considered accurate.
We see that the proposed approach leads to significant improvements for all the tested number of microphones.
}

\begin{figure}
 \begin{center}
 \subfloat[]
 {
\includegraphics[width=0.9\linewidth]{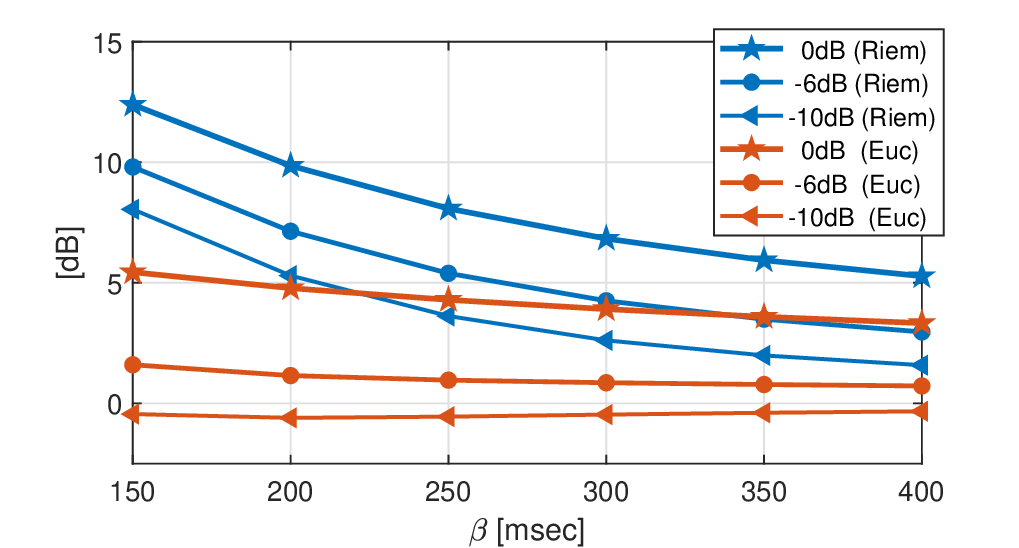}
}\\
\subfloat[]
{
\includegraphics[width=0.9\linewidth]{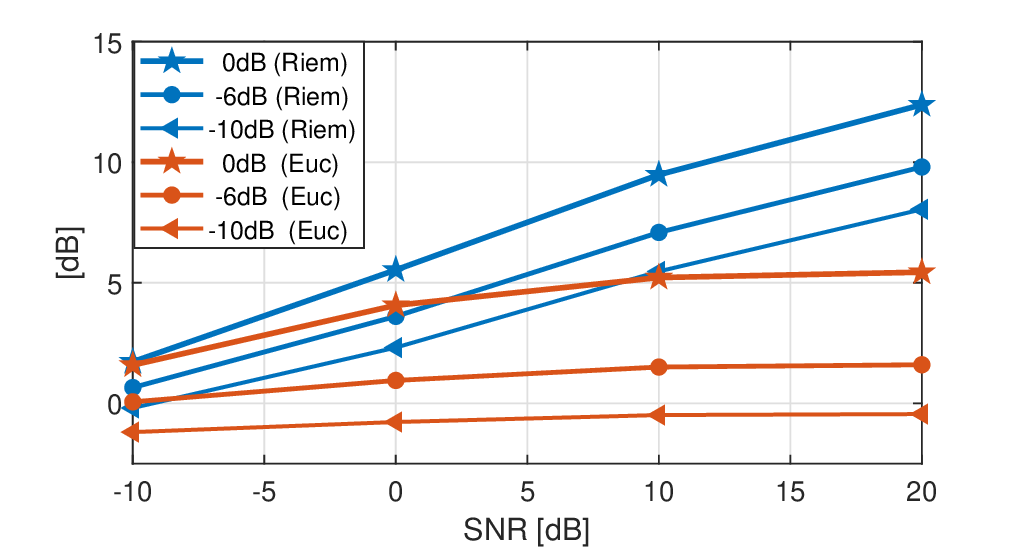}
}
\caption{The mean output SIR as a function of (a) the reverberation times and (b) the SNR, for $2$ interference sources. The Riemannian DS appears in blue, whereas the Euclidean DS appears in red. Several input SIR values are presented: $0\text{dB}$ (asterisks), $-6\text{dB}$ (circle), and $-10\text{dB}$ (triangle). }
\label{Fig: Mean Min Sir many beta 40dB SNR}
\end{center}
\end{figure}

Next, we examine the sensitivity of the proposed approach to the SNR and the reverberation time. 
We repeat the setting of the two interferences, as described in the first experiment. The results are presented in Figure \ref{Fig: Mean Min Sir many beta 40dB SNR}. At the top, the mean output SIR for the DS method is presented as a function of the reverberation time for a fixed  SNR of ${2}0\text{dB}$. Several input SIR values are shown: $0\text{dB}$ (asterisks), $-6\text{dB}$ (circle), and $-10\text{dB}$ (triangle). The results for the Riemannian and Euclidean DS appear in blue and red, respectively. The bottom figure is the same as the top, only for different SNR values for a fixed reverberation time of $\beta=150\text{ms}$. %
We see that the smaller the reverberation time is, the higher the output SIR is for the Riemannian DS method. In contrast, the Euclidean DS is less affected by the reverberation time, resulting in relatively low output SIRs. For all values of reverberation times, the Riemannian DS results in higher output SIR than its Euclidean counterpart.
From the bottom figure, we see that the SNR has a large impact on the performance of the Riemannian DS; the higher the SNR is, the higher the output SIR becomes. Conversely, the Euclidean DS is moderately affected by the SNR, resulting in much lower output SIR values.
A possible explanation is that the main phenomenon limiting the performance of the Euclidean approach is the existence of interference sources. 
In addition, as Proposition \ref{prop: derivatives of Riemannian SIR} predicts, the sensitivity of the Riemannian DS to the SNR is higher than the Euclidean DS, and the higher the SNR is, the larger the gap in the performance between the Riemannian and the Euclidean approaches.

We examine the performance of the MVDR beamformer, given by (\ref{eq: MVDR spectrum}). 
The MVDR beamformer is popular when interference sources are present, thanks to its distortionless response in the direction of the desired source, and its typical narrow beams. However, in our setting, since the desired source is accompanied by interference sources, and the directions to all the sources are unknown, the DoA estimation of the desired source using the MVDR beamformer is outperformed by the DS and the SbSp methods.
We also examine the case of $2$ desired sources.
The results show similar trends and appear in Appendix \ref{Apppendix: additional experimental results}, due to space considerations.

\begin{figure}
 \begin{center}
 \subfloat[]
 {
\includegraphics[width=0.48\linewidth]{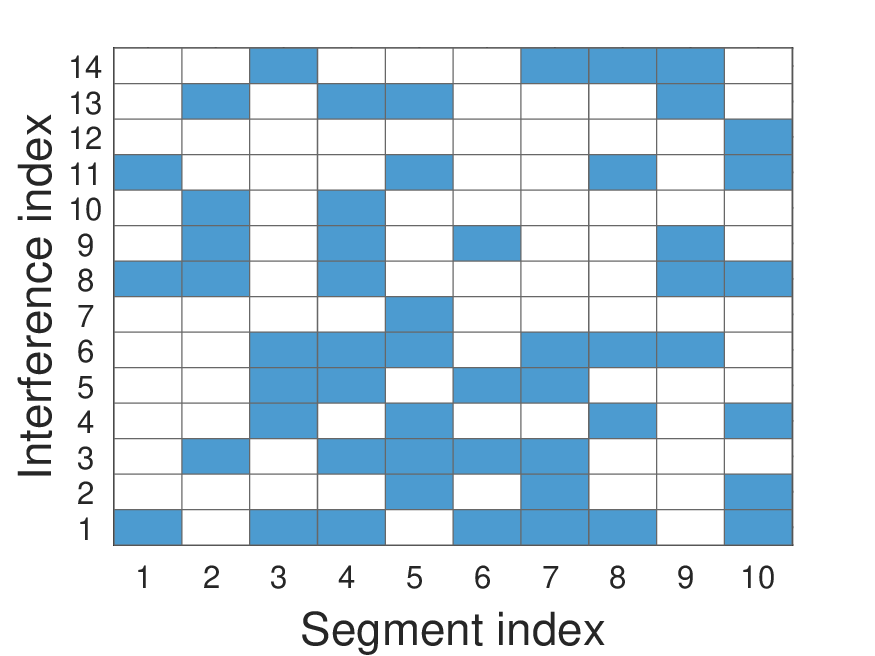}
}
\subfloat[]
{
\includegraphics[width=0.5\linewidth]{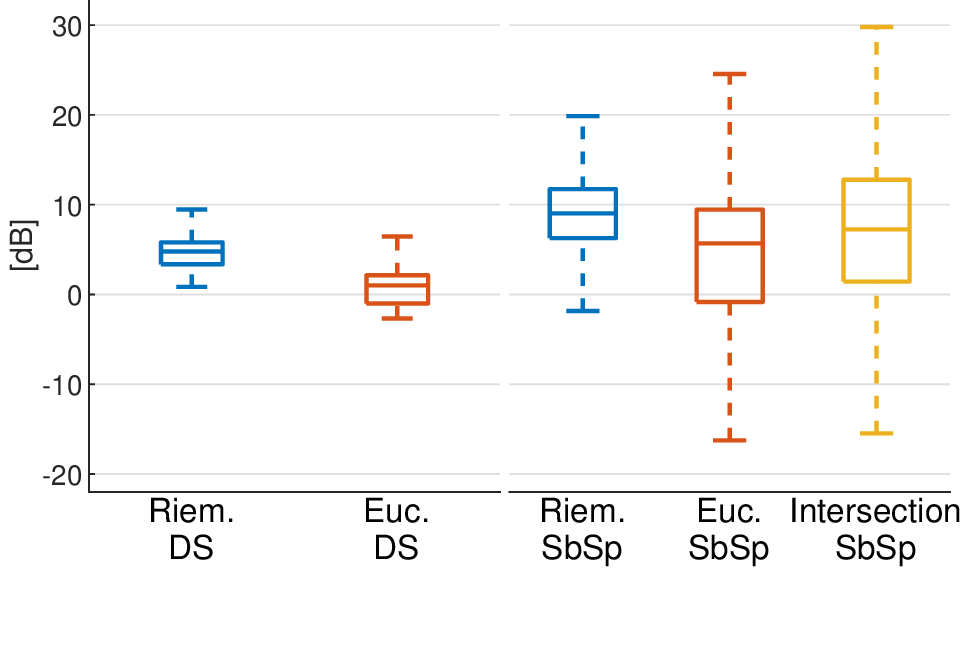}
}
\caption{(a) Activation map for the $14$ interference sources during the observed $10$ time segments (blue indicates `active'). (b) Output SIR of the different beamformers. The input SIR of each interference is $-6\text{dB}$.}
\label{Fig: Activatio map 14 interferences}
\end{center}
\end{figure}

In the second experiment, we examine a multiple interference setting, by considering $N_{\text{I}} = 14$ interference sources. We note that the number of interference sources is larger than the number of microphones, $N_{\text{I}} > M = 12$, which typically limits the number of interference sources that can be accommodated (e.g. see \cite{markovich2009multichannel}). Furthermore, assumptions \ref{assump: uncorrelated ATFs of target and interferences} and \ref{assump: uncorrelated ATFs between interferences} implicitly restrict the number of interference sources to be bounded by $M-N_{\text{D}} = 11$. This limitation is only for the analysis, and, in practice, improved results are obtained even for a larger number of interference sources. The signal contains $10$ segments, demonstrating the number of segments could be smaller than the number of interference sources. The duration of the emitted signal is $10.24\text{s}$. 
The input SIR for each interference is $-6\text{dB}$. Each interference has a $30\%$ probability of being active at each segment. The position of all the sources is set at random on the arc, as described in the first experiment. The activation map of the interference sources at the first Monte Carlo iteration appears in Figure \ref{Fig: Activatio map 14 interferences}(a), where light blue indicates `active'. On average, ${4.2}$ interference sources are active at the same time at each segment. An interference source that is partially active during a segment is considered active during the entire segment (a worst case). Note that there are interference sources active continuously during more than one segment, so their activation is not necessarily related to the division of the received signal into segments. 
Additionally, there exists no segment in which only one source is active. The output SIRs are presented in Figure \ref{Fig: Activatio map 14 interferences}(b). The Riemannian DS and SbSp methods appear in blue, the Euclidean DS and SbSp methods are in red, and the intersection is in orange. It can be seen that the Riemannian approach is superior to the Euclidean one, resulting in higher output SIRs.  
   {We note that Figure \ref{Fig: Activatio map 14 interferences}(b) serves as an example of an increased number of segments in comparison to Figure 3. Since there are on average $4.2$ active interference sources at every segment, the results are not improved as the number of segments is increased. } 
   
{We remark that the empirical results of the proposed approach applied to the typically-used beamformers are not sensitive to the number of samples used for the computation of the correlation matrix. Since, in the experiments, additional samples also include samples from at least one interference source, adding samples increases the effect of the interference sources, resulting in weak dependency on the number of samples.} 

\begin{figure}
 \begin{center}
 \subfloat[]
 {
\includegraphics[width=0.49\linewidth]{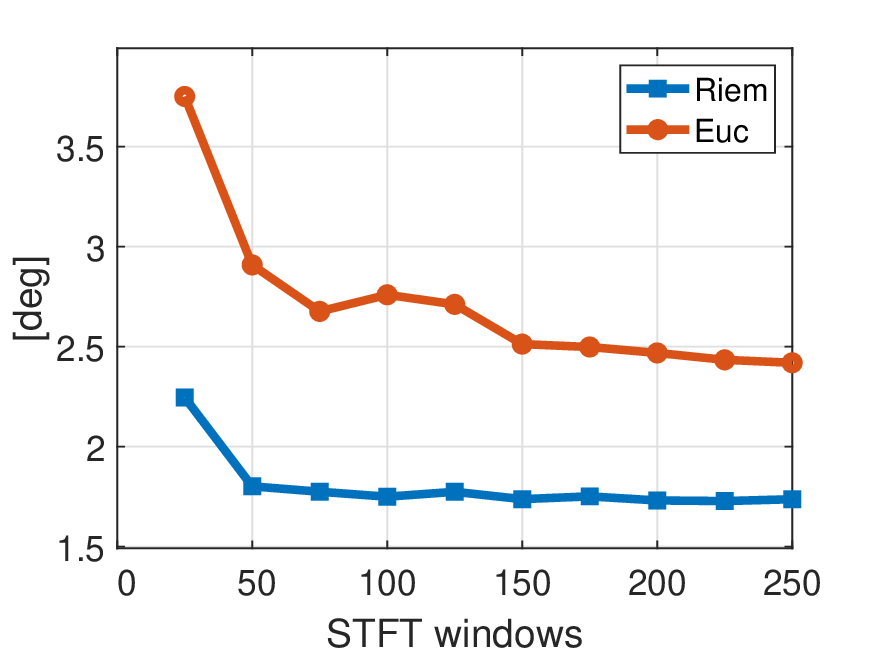}
}
\subfloat[]
{
\includegraphics[width=0.49\linewidth]{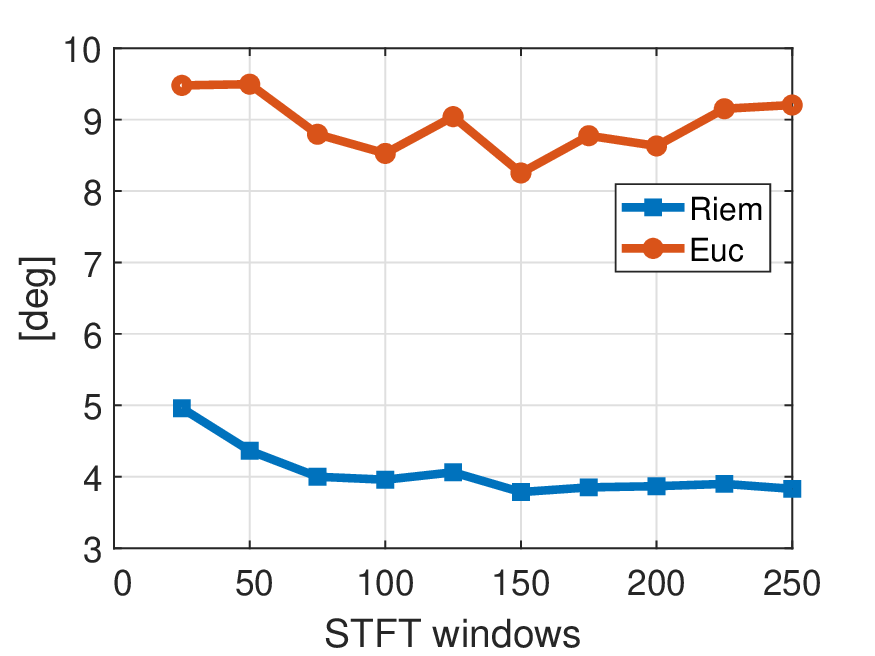}
}
\caption{{The RMSE of the Bayesian learning method for $6$ desired sources with and without the proposed approach in blue and red, respectively, for (a) SIR $-6$dB and (b) SIR $-10$dB.}}
\label{fig: RMSE bayesian vs snapshots 2}
\end{center}
\end{figure}

{Next, we evaluate the performance of the proposed approach applied to the Bayesian learning method proposed in \cite{hu2016source} for sparse signal recovery for DoA estimation. This is done by replacing the typically-used (Euclidean) sample correlation matrix with the Riemannian mean of sample correlation matrices computed in short-time segments. We repeat the same experiment as in \cite{hu2016source} with $6$ desired sources, only we consider two disjointly active interference sources with SIR $-6$dB and $-10$dB. The SNR is $20$dB. The number of desired sources is unknown. 
We evaluated the performance using $3000$ experiments, where at each experiment the direction of the interference sources is generated uniformly at random between $[-50^\circ,50^\circ]$.
Figure \ref{fig: RMSE bayesian vs snapshots 2} presents the RMSE for a different number of STFT windows, which are equivalent to the number of samples for the sample correlation matrix computation of each segment (there are two segments in the experiment). We see that applying the proposed Riemannian approach leads to improved performance for all the tested number of samples.}

\section{Conclusion}
\label{sec: Conclusion}
We present a Riemannian approach for the design of beamformers {and DoA estimation methods} for interference rejection in reverberant environments. {Specifically, the Riemannian mean is incorporated instead of the sample correlation matrix} for DoA estimation of the desired source, as it {inherently} rejects the interference sources. We analytically show that the DS beamformer, based on the Riemannian geometry of the HPD manifold, results in a higher output SIR than the typical DS beamformer, which implicitly considers the Euclidean geometry. We extend our approach to other beamformers, such as subspace-based beamformers and the MVDR, {as well as a Bayesian learning method}, experimentally demonstrating superior output SIR and better DoA estimations in comparison to their Euclidean counterparts.

%

%

%

%



{\appendices
\section{The Intersection Beamformer}
\label{Apppendix: The Intersection Method}
Another beamformer we examine is based on the observation that the desired signal subspace is the intersection of subspaces spanned by eigenvectors of the correlation matrix of the different segments.
From each segment, $i$, we extract the desired signal subspace from the {sample} correlation matrix, $\sGamma_i$, to obtain $\mV(\sGamma_i)$ whose columns are the $N_\text{D}$ leading eigenvectors.
The projection matrix onto the signal subspace of $\sGamma_i$ is computed as follows:
\begin{equation}
\mP_{\text{sig}}(\sGamma_i) = \mV(\sGamma_i)\left(\mV^H(\sGamma_i)\mV(\sGamma_i)\right)^{-1}\mV^H(\sGamma_i).    
\end{equation}
Since each desired source is active during all the segments, its ATF, $\vh^d_j$, is an eigenvector of $\mV(\sGamma_i)$. Consequently it is also an eigenvector of $\mP_{\text{sig}}(\sGamma_i)$ for all $i$, with an eigenvalue $1$, i.e. $\mP_{\text{sig}}(\sGamma_i)\vh^d_j = \vh^d_j$.
As a result, it is an eigenvector of $\mP_{\text{sig}}  =\prod_{i=1}^{L_{\text{s}}} \mP_{\text{sig}}(\sGamma_i)$ with eigenvalue $1$ as well since $\mP_{\text{sig}} \vh^d_j =\left(\prod_{i=1}^{L_{\text{s}}} \mP_{\text{sig}}(\sGamma_i)\right)\vh^d_j = \vh^d_j$.
We consider the leading $N_{\text{D}}$ eigenvectors of $\mP_{\text{sig}}$
and form the matrix $\mP_{\text{sig},N_{\text{D}}}\in\sC^{M\times N_{\text{D}}}$, whose columns are the eigenvectors. 
The intersection {spectrum} is defined as follows:
\begin{equation}
    \mP_{\text{Intersect}}(\theta) = \vd^H(\theta)\mP_{\text{sig},N_{\text{D}}}\mP_{\text{sig},N_{\text{D}}}^H\vd(\theta)   .
\end{equation}
We note that in addition to the dimension of the signal space of the matrix $\mP_{\text{sig}}$, the intersection method requires the estimation of the dimension of the signal space of the correlation matrix for each segment (in order to compute $\mV(\sGamma_i)$).

\section{Additional experimental results}
\label{Apppendix: additional experimental results}
We repeat the setting of the first experiment, only with an additional desired source, located at $(2\text{m}, 3.5\text{m}, 2.5\text{m})$. %
Figure \ref{Fig: Mean SIR 2 interferences 2 sources SIR -6 -10} is the same as Figure \ref{Fig: Mean SIR Directivity 2 interferences ML}, but the mean output SIR is taken over the two interferences and the two desired sources. The left figure presents results for input SIR of $-10\text{dB}$, and the right presents results for input SIR of $-6\text{dB}$. We see that the Riemannian approach is superior to the Euclidean one, resulting in higher SIRs. Also, the intersection method is more sensitive to the input SIR, resulting in higher output SIR for $-6\text{dB}$. Similar to the setting of one desired source, the Riemannian SbSp method is superior to the Riemannian DS method, as opposed to the Euclidean approach, for which they are on par. This is a consequence of the higher interference attenuation of the Riemannian approach, resulting in a better estimate of the signal space dimension.

\begin{figure}
 \begin{center}
 \subfloat[]
 {
\includegraphics[width=0.48\linewidth]{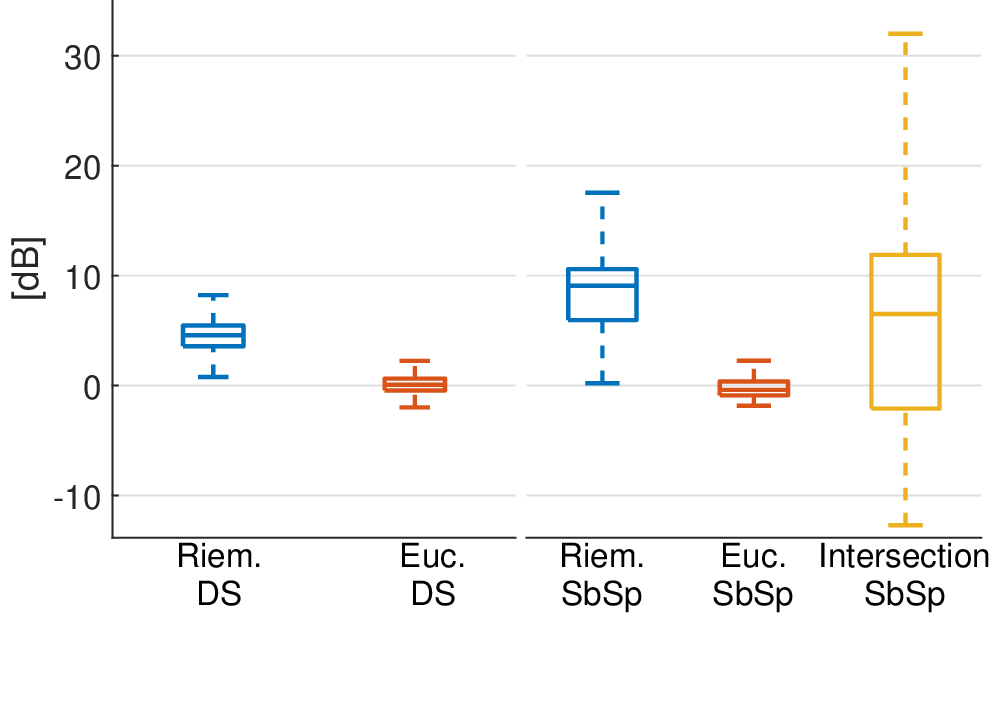}
}
\subfloat[]
{
\includegraphics[width=0.48\linewidth]{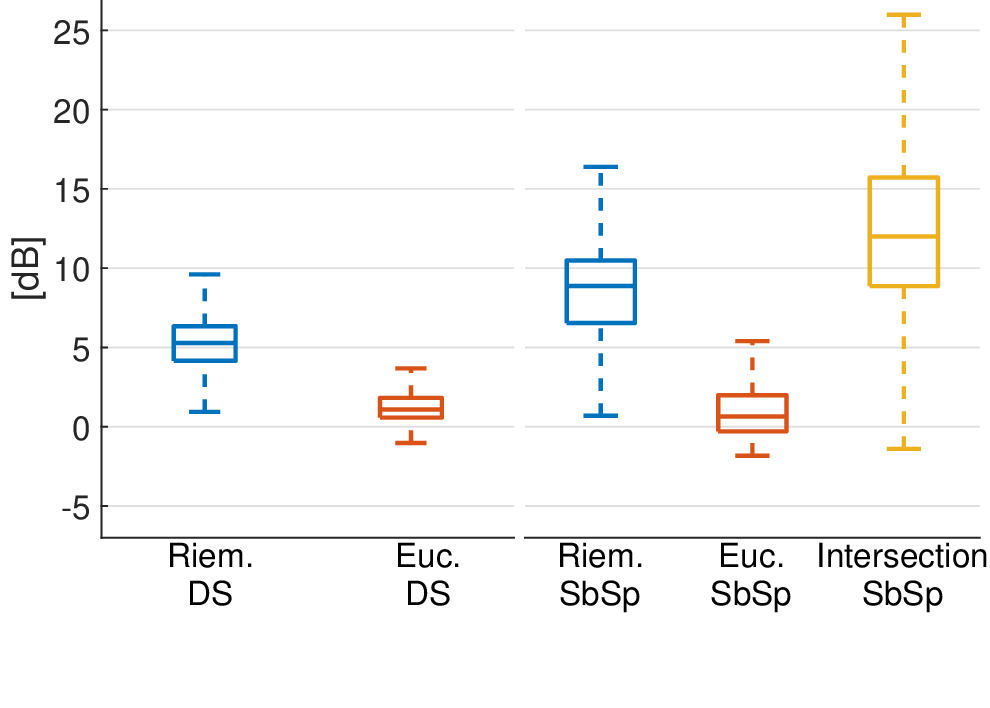}
}
\caption{Output SIR of the different beamformers for two desired sources and two interference sources. (a) The input SIR is $-10$dB. (b) The input SIR is $-6$dB.}
\label{Fig: Mean SIR 2 interferences 2 sources SIR -6 -10}
\end{center}
\end{figure}

Next, we examine the performance of the Riemannian MVDR, and the Euclidean MVDR beamformers, given by (\ref{eq: MVDR spectrum}),
for $\sGamma_{\text{R}}$ and $\sGamma_{\text{E}}$, respectively.
Figure \ref{Fig: Mean SIR 2 interferences MVDR} presents the results. It is the same as Figure \ref{Fig: Mean SIR Directivity 2 interferences ML}, only for the MVDR beamformer. We see that the Riemannian approach is superior to the Euclidean one, however, with a slight advantage.

\begin{figure}
 \begin{center}
\includegraphics[width=0.95\linewidth]{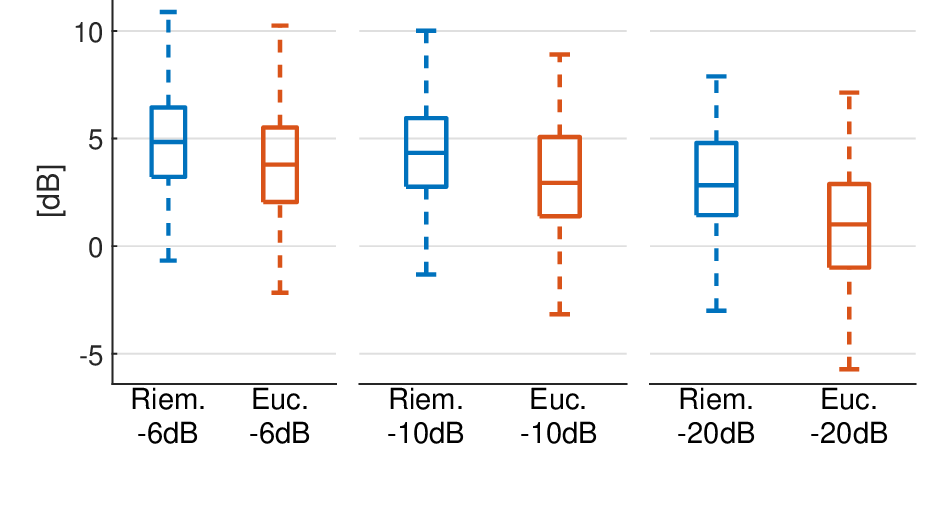}
\caption{The mean output SIR for the Riemannian and the Euclidean MVDR beamformers in the presence of two interference sources. The x-axis indicates the input SIR, and the y-axis indicates the output SIR.}
\label{Fig: Mean SIR 2 interferences MVDR}
\end{center}
\end{figure}
{We repeat the experiment with speech signals from the TIMIT dataset. For each source, the speaker and the time interval are chosen uniformly at random. Figure \ref{fig: DoA speech sig Fig5} presents the results. We see that the proposed approach leads to improved results in comparison to the typically-used beamformers also for speech signals.}
\begin{figure}
 \begin{center}
 \subfloat[]
 {
\includegraphics[width=0.49\linewidth]{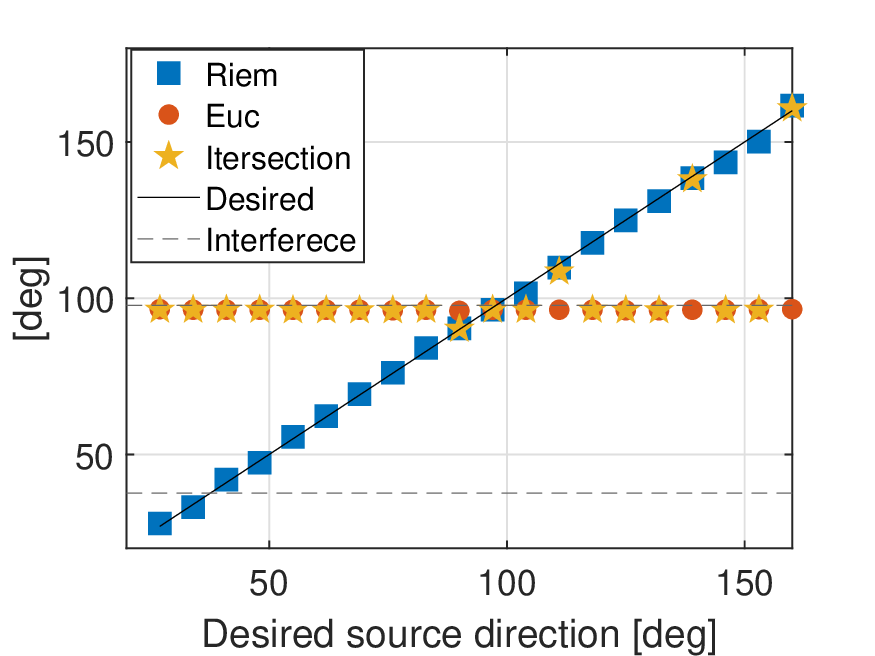}
}
\subfloat[]
{
\includegraphics[width=0.49\linewidth]{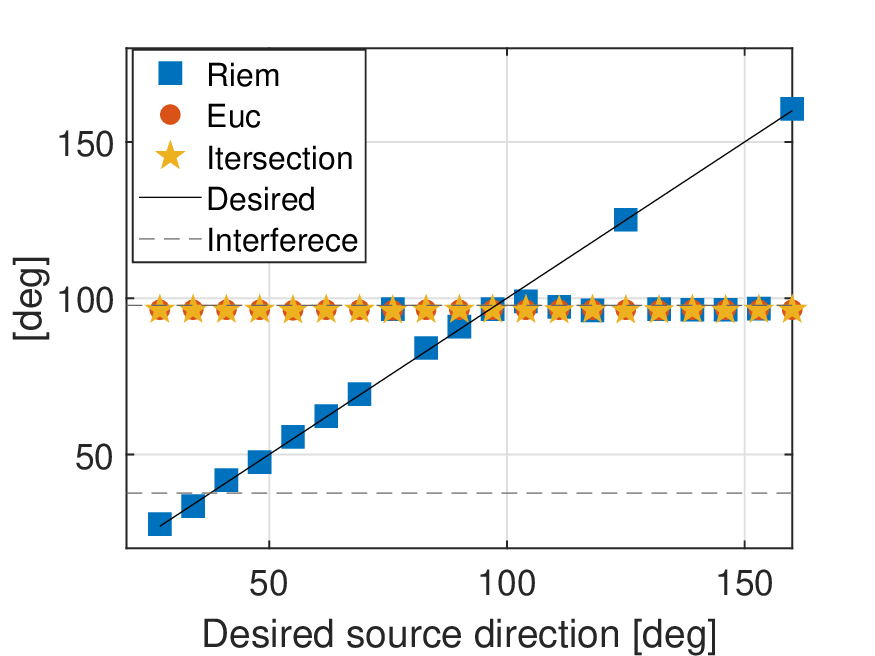}
}
\caption{{Estimation of the DoA to the desired source for speech signals where (a) the input SIR is -6dB, and (b) the input SIR is $-10$dB.}}
\label{fig: DoA speech sig Fig5}
\end{center}
\end{figure}

\section{Extension to a Streaming Data Setting}

\label{subsec: Extension to a Streaming Setting}

In a streaming data setting, we do not have access in advance to the entire signal, and the direction estimation is updated as more samples become available. 
Therefore, in this setting, we cannot compute the Riemannian mean using (\ref{eq: Riem mean as argmin}). To circumvent this, we turn to the estimator of the Riemannian mean proposed in \cite{ho2013recursive,cheng2016recursive}, which is updated after every received segment. 

Each segment, $i$, is processed separately using $L_{\text{w}}$ STFT windows, and an estimation of the correlation matrix, $\sGamma_i$, is computed using (\ref{eq: general correlation matrices}). 
Next, the estimation of the Riemannian mean, denoted as $\hat{\mR}_i$, which is the adaptive counterpart of $\mathbf{\Gamma}_{\text{R}}$, is updated using the following update step:
\begin{equation}
    \hat{\mR}_{i} = \hat{\mR}_{i-1}^{\frac{1}{2}}(\hat{\mR}_{i-1}^{-\frac{1}{2}}\sGamma_{i}\hat{\mR}_{i-1}^{-\frac{1}{2}})^{\frac{1}{i}}\hat{\mR}_{i-1}^{\frac{1}{2}}.
\end{equation}

The {SRP} of the DS beamformer is computed using (\ref{eq: ML Spectrum}) with $\hat{\mR}_i$ as an estimate of $\sGamma_{\text{R}}$, and the direction to the desired source is set using (\ref{eq: theta hat from P_ML}). The algorithm is described in Algorithm \ref{Alg: K points online}.

We note that the current correlation matrix estimate has a full rank if $L_{\text{w}} > M$. This implies a latency of at least $M$ times the duration of the STFT window. In case the STFT is performed with overlap this latency is reduced accordingly.

\begin{algorithm}
\caption{Streaming DoA estimation in the presence of multiple interferences}
\label{Alg: K points online}
\textbf{Input:} The current result of the STFT window of the received signal\\
\textbf{Output:} The estimated direction of the desired source $\hat{\theta}$\\
\begin{algorithmic}[1]
\STATE Set $\hat{\mR}_0 = \mI$
\STATE Repeat %
\begin{enumerate}
    \item \label{step: Accumulate L_w stft windows}
    Accumulate $L_{\text{w}}$ STFT windows (that form a segment)
    \item Compute its {sample} correlation matrix, $\sGamma_i$, using (\ref{eq: general correlation matrices})
    \item \label{step: R_i hat}
    Compute $\hat{\mR}_{i} = \hat{\mR}_{i-1}^{\frac{1}{2}}(\hat{\mR}_{i-1}^{-\frac{1}{2}}\sGamma_{i}\hat{\mR}_{i-1}^{-\frac{1}{2}})^{\frac{1}{i}}\hat{\mR}_{i-1}^{\frac{1}{2}}$
    \item Compute $P_\text{DS}(\theta;\hat{\mR}_{i})$  using (\ref{eq: ML Spectrum})
    \item Return $\hat{\theta}=\argmax_\theta P_\text{DS}(\theta;\hat{\mR}_{i})$
\end{enumerate}
\end{algorithmic}
\end{algorithm}

As for the Euclidean alternative, its estimator is updated in finer granularity at every STFT window, $l$. The estimator at the $l$th update is denoted by $\mE_l$ and is computed as follows:
\begin{equation}
\label{eq: online update of Euclidean mean}
    \mE_l = \frac{n-1}{n}\mE_{l-1} + \frac{1}{n}\vz(l)\vz^H(l) .
\end{equation}
Setting $L_{\text{w}}=1$ in step 2.\ref{step: Accumulate L_w stft windows} in Algorithm \ref{Alg: K points online}, and substituting (\ref{eq: online update of Euclidean mean}) in step 2.\ref{step: R_i hat} in Algorithm \ref{Alg: K points online}, results in the streaming version of the Euclidean counterpart.

\section{On the Particular Choice of the Riemannian Metric}
\label{Appendix: On the Particular Choice of the Riemannian Metric}
In this work, we consider the Affine Invariant metric (also called Fisher Information metric)  \cite{pennec2006riemannian}. 
Another commonly-used metric in the space of HPD matrices is the Log-Euclidean metric \cite{arsigny2007geometric}, which could be viewed as a computationally efficient local approximation of the Affine Invariant metric. The induced Log-Euclidean distance is given by:
\begin{equation}
    d_{\text{LE}}^2(\mathbf{\Gamma}_1,\mathbf{\Gamma}_2) = 
    \|\log (\mathbf{\Gamma}_1)-\log(\mathbf{\Gamma}_2) \|_{F}^2.
\end{equation}
In the context of this work, the Affine Invariant metric is advantageous over the Log Euclidean because it better enhances the desired source subspace relative to the interference and noise subspace, as we show next. 
We remark that the derivation is similar to the derivation in Section \ref{sec: Analysis}, where we demonstrate the advantage of the Affine Invariant metric over Euclidean geometry.
First, we note that the following holds \cite{congedo2015approximate}:

\begin{equation}
\label{eq: tr(Gamma_R) < tr(Gamma_LE) }
    \text{tr}(\mathbf{\Gamma}_{\text{R}}) \le
    \text{tr}(\mathbf{\Gamma}_{\text{LE}}),
\end{equation}
where $\mathbf{\Gamma}_{\text{LE}}$ denotes the Riemannian mean based on the Log-Euclidean metric.
Second, the ATF to the desired source, $\vh_0$, is a common eigenvector of all the correlation matrices per segment.
Consequently, according to Lemma \ref{lemma: common eigenvalue equality} below, the mean correlation matrices based on both metrics have the same eigenvalue associated with the common eigenvector $\vh_0$.
More specifically, denoting $\lambda_0(\mathbf{\Gamma}) \equiv \frac{\vh_0^H\mathbf{\Gamma}\vh_0}{\|\vh_0\|^2}$, we get $\lambda_0(\mathbf{\Gamma}_{\text{R}})=\lambda_0(\mathbf{\Gamma}_{\text{LE}})$.

\begin{lemma}
\label{lemma: common eigenvalue equality}
Let $\{\mathbf{\Gamma}_j\}_j$ be the set of HPD matrices, having a common eigenvalue $\lambda_0$, associated with the common eigenvector $\vh_0$. Then 
\begin{equation}
    \begin{split}
        \vh_0^H \mathbf{\Gamma}_{\text{R}} \vh_0 =
        \vh_0^H \mathbf{\Gamma}_{\text{LE}} \vh_0 = 
        \vh_0^H \mathbf{\Gamma}_{\text{E}} \vh_0 = 
      \|\vh_0\|^2 \lambda_0.
    \end{split}
\end{equation}
where $\mathbf{\Gamma}_{\text{R}}$, $\mathbf{\Gamma}_{\text{LE}}$, $\mathbf{\Gamma}_{\text{E}}$ are the Riemannian means based on the Affine Invariant and the Log-Euclidean metrics, and the Euclidean mean.
\end{lemma}

The rest of the eigenvectors of the correlation matrices span the interference and noise subspace. We recall that %
$\lambda_i(\mathbf{\Gamma})$ is the $i$th eigenvalue, so we get
\begin{equation}
\label{eq: subspace(Gamma_R) > subspace(Gamma_E)}
    \frac{\lambda_0(\mathbf{\Gamma}_{\text{R}})}{\sum_{i=1}^{M-1}\lambda_i(\mathbf{\Gamma}_{\text{R}})} \ge
    \frac{\lambda_0(\mathbf{\Gamma}_{\text{LE}})}{\sum_{i=1}^{M-1}\lambda_i(\mathbf{\Gamma}_{\text{LE}})},
\end{equation}
which follows from Lemma \ref{lemma: common eigenvalue equality}, %
and (\ref{eq: tr(Gamma_R) < tr(Gamma_LE) }).
We see that the Riemannian mean induced by the Affine Invariant metric captures better the desired signal subspace in comparison to the Riemannian mean induced by the Log Euclidean metric and the Euclidean mean (according to (\ref{eq: subspace R > subspace E})), entailing an advantage in SbSp methods, for example.

}

\bibliographystyle{IEEEtran}
\bibliography{Refs}

\clearpage

\onecolumn
     
\section*{\LARGE Supplementary Material: On Interference-Rejection Using Riemannian Geometry for Direction of Arrival Estimation}

\normalsize
\setcounter{equation}{57}
\setcounter{proposition}{4}
\setcounter{lemma}{1}

\vspace*{30px}

\section*{Proofs of Theoretical Results}
\label{Appendix: Proofs for the Theoretic Results}
A key observation in our analysis is that although there is no explicit expression for the Riemannian mean, in general, under assumptions \ref*{assump: uncorrelated ATFs of target and interferences} and \ref*{assump: uncorrelated ATFs between interferences}, the correlation matrices $\{\mathbf{\Gamma}_l\}$ commute, and the Riemannian mean has an explicit form, according to Proposition \ref{prop: Riemannian mean for commuting matrices}.
We note that in practice we use Algorithm \ref{Alg: Riemannian mean interations} to compute the Riemannian mean because the assumptions do not strictly hold.

For completeness, we prove Proposition \ref{prop: Riemannian mean for commuting matrices}, which is a property of Karcher's mean \cite{lim2012matrix}.

\begin{proposition}
\label{prop: Riemannian mean for commuting matrices}
The Riemannian mean of $K$ commuting HPD matrices, $\{\mathbf{\Gamma}_l\}_{l=1}^K$ is given by
\begin{equation}
    \mathbf{\Gamma}_{\text{R}} = 
    \prod_{l=1}^K\mathbf{\Gamma}_l^{\frac{1}{K}}.
\end{equation}
\begin{proof}
The Karcher mean, $\mathbf{\Gamma}_{\text{R}}$, is the solution of the Karcher equation
\cite{karcher1977riemannian,bhatia2009positive,moakher2005differential,lim2012matrix}
\begin{equation}
    \sum_{l=1}^K\log(\mathbf{\Gamma}_{\text{R}}^{\frac{1}{2}}\mathbf{\Gamma}_l^{-1}\mathbf{\Gamma}_{\text{R}}^{\frac{1}{2}}) = \mathbf{0}
\end{equation}
We set $\mathbf{\Gamma}_{\text{R}}= \prod_{l=1}^K\mathbf{\Gamma}_l^{\frac{1}{K}}$, and use the fact the matrices commute:
\begin{equation}
\begin{split}
    \sum_{l=1}^K\log(\mathbf{\Gamma}_{\text{R}}^{\frac{1}{2}}\mathbf{\Gamma}_l^{-1}\mathbf{\Gamma}_{\text{R}}^{\frac{1}{2}})   
    &=
    \sum_{l=1}^K\log(\mathbf{\Gamma}_{\text{R}}\mathbf{\Gamma}_l^{-1})  \\
    & =
    \sum_{l=1}^K\log\left(\mathbf{\Gamma}_l^{-1}\prod_{l=1}^K\mathbf{\Gamma}_l^{\frac{1}{K}}\right)  \\
    & =
    \log\prod_{l=1}^K\left(\mathbf{\Gamma}_l^{-1}\prod_{l=1}^K\mathbf{\Gamma}_l^{\frac{1}{K}}\right)  \\
    & =
    \log\mI \\
    & = \mathbf{0}.
\end{split}
\end{equation}
\end{proof}
\end{proposition}

\subsection{Proof of Proposition \ref*{prop: SIR and comparison}}
\begin{proof}
The proof relies on Lemma \ref*{lemma: Riemannian Euclidean mean}, which we prove first.
\setcounter{lemma}{0}
\begin{lemma}
The Riemannian or the Euclidean mean of the population correlation matrices of the segments (\ref{eq: corr matrix}) over the entire interval can be written in the same parametric form as:
\begin{equation}
\begin{split}
    \mathbf{\Gamma} = 
     \sigma_0^2\vh_0\vh_0^H + 
    \sum_{j=1}^{N_{\text{I}}} \mu^2_j \vh_j\vh_j^H +
    \sigma_v^2\mI.
\end{split}
\end{equation}
The Riemannian mean $\mathbf{\Gamma}_{\text{R}}$ is obtained by setting the parameters $\mu_j$ to 
\begin{equation}
\mu_j^2 = 
\frac{(\sigma_j^2\Vert \vh_j\Vert^2 + \sigma_v^2)^{\tau_j}(\sigma_v^2)^{1-\tau_j}-\sigma_v^2}{\Vert \vh_j\Vert^2},
\end{equation}
and the Euclidean mean $\mathbf{\Gamma}_{\text{E}}$ is obtained by setting
\begin{equation}
\mu_j^2 =
 \sigma_j^2 \tau_j.
\end{equation}
\begin{proof}
We start by expressing the correlation matrix of the $i$th segment:
\begin{equation}
\label{eq: Gamma_l}
    \mathbf{\Gamma}_i = 
    \sigma_0^2\vh_0\vh_0^H + 
    \sum_{j\in \mathcal{J}_i} \sigma_j^2\vh_j\vh_j^H +
    \sigma_v^2\mI,
\end{equation}
where $\mathcal{J}_i$ is the set of indices of the active interference sources during the $i$th segment.
The vectors $\vh_0$, and $\{\vh_j\}_{j=1}^{N_{\text{I}}}$ are all orthogonal, thus they are eigenvectors of $\mathbf{\Gamma}_i$, $\forall i=1,...,L_{\text{s}}$.

We denote by upper tilde the normalized version of a vector, i.e. $\tilde{\vh}=\frac{\vh}{\|\vh\|}$. So,
\begin{equation}
    \begin{split}
        \tilde{\vh}_0^H\mathbf{\Gamma}_i\tilde{\vh}_0 &=
        \sigma_0^2\|\vh_0\|^2 + \sigma_v^2 
         \;\;\; \forall i \\
        \tilde{\vh}_j^H\mathbf{\Gamma}_i\tilde{\vh}_j &=
        \sigma_j^2\|\vh_j\|^2 + \sigma_v^2 
         \;\;\; i\in\mathcal{L}_j \\
        \tilde{\vh}_j^H\mathbf{\Gamma}_l\tilde{\vh}_j &=
        \sigma_v^2 \;\;\; i\notin\mathcal{L}_j
    \end{split}
\end{equation}

The correlation matrices for each segment, $\{\mathbf{\Gamma}_i\}_{i=1}^{L_{\text{s}}}$, commute with each other because they share their eigenvectors: $\vh_0,\{\vh_j\}_{j=1}^{L_I}$, and $\{\vv_l\}_{l=1}^{M-N_{\text{I}}-1}$, where $\{\vv_j\}_{j=1}^{M-N_{\text{I}}-1}$ are the eigenvectors spanning the common noise subspace of all the matrices.

Following Proposition \ref{prop: Riemannian mean for commuting matrices}, it holds true that
\begin{equation}
     \mathbf{\Gamma}_{\text{R}} = 
    \prod_{j=1}^{L_{\text{s}}}\mathbf{\Gamma}_j^{\frac{1}{L_{\text{s}}}}.
\end{equation}
We compose $\mathbf{\Gamma}_{\text{R}}$ using a transformation for the eigenvalues for each $\mathbf{\Gamma}_i$:
\begin{equation}
\begin{split}
    \tilde{\vh}_j^H\mathbf{\Gamma}_{\text{R}}\tilde{\vh}_j &=
    ( \sigma_j^2\|\vh_j\|^2+\sigma_v^2)^{\tau_j}(\sigma_v^2)^{1-\tau_j}, \;\;\; \forall j \\
    \tilde{\vh}_0^H\mathbf{\Gamma}_{\text{R}}\tilde{\vh}_0 &=  \sigma_0^2\|\vh_0\|^2+\sigma_v^2,
\end{split}
\end{equation}
and the rest of the eigenvectors have an eigenvalue of $\sigma_v^2$, with multiplicity of $M-N_{\text{I}}-1$.
The matrix $\mathbf{\Gamma}_{\text{R}}$ meets all the requirements.

The proof for the Euclidean mean follows from its definition and (\ref{eq: Gamma_l}).
\end{proof}
\end{lemma}

We compute the output SIR for a general matrix with a similar structure as in Lemma \ref*{lemma: Riemannian Euclidean mean}. The correlation matrix is:
\begin{equation}
    \begin{split}
        \mathbf{\Gamma} = 
        \sigma_0^2\vh_0^H\vh_0 + 
        \sum_{l=1}^{N_{\text{I}}} \mu_l^2 \vh_l^H\vh_l +
    \sigma_v^2\mI,
    \end{split}
\end{equation}
and the SIR becomes:
\begin{equation}
\label{eq: SIR expression from proof prop 1}
    \begin{split}
        &\text{SIR}_j(\mathbf{\Gamma})\\  
        & =
        \frac{
        \vd_0^H\left(
        \sigma_0^2\vh_0^H\vh_0 + 
        \sum_{l=1}^{N_{\text{I}}} \mu_l^2 \vh_l^H\vh_l +
    \sigma_v^2\mI
    \right)\vd_0
    }
    {
     \vd_j^H\left(
        \sigma_0^2\vh_0^H\vh_0 + 
        \sum_{l=1}^{N_{\text{I}}} \mu_l^2 \vh_l^H\vh_l +
    \sigma_v^2\mI
    \right)\vd_j 
    }                          \\
    &=
    \frac{
    \sigma_0^2|\langle\vd_0^H,\vh_0\rangle|^2 + 
        \sum_{l=1}^{N_{\text{I}}} \mu_l^2 |\langle\vd_0^H,\vh_l\rangle|^2 +
    \sigma_v^2 M 
    }
    {
    \sigma_0^2|\langle\vd_j^H,\vh_0\rangle|^2 + 
        \sum_{l=1}^{N_{\text{I}}} \mu_l^2 |\langle\vd_j^H,\vh_l\rangle|^2 +
    \sigma_v^2 M
    }                  \\
   & =
  \frac{
  \sigma_0^2M\|\vh_0\|^2\kappa+ 
        \sum_{l=1}^{N_{\text{I}}} \mu_l^2 M \|\vh_l\|^2 \rho +
    \sigma_v^2 M 
    }
    {  \splitdfrac{   
    \sigma_0^2 M \|\vh_0\|^2\rho + 
        \sum_{l=1,l\neq j}^{N_{\text{I}}} \mu_l^2 M \|\vh_l\|^2\rho  
        +\mu_j^2 M \|\vh_j\|^2\kappa}{ + 
    \sigma_v^2 M }
    }                      \\
     & =
    1     +
    \frac{
    (\sigma_0^2\|\vh_0\|^2-\mu_j^2\|\vh_j\|^2)\kappa+ 
        (\mu_j^2\|\vh_j\|^2-\sigma_0^2\|\vh_0\|^2) \rho
    }
    {
    \sigma_0^2\|\vh_0\|^2\rho + 
        \sum_{l=1,l\neq j}^{N_{\text{I}}} \mu_l^2 \|\vh_l\|^2\rho + \mu_j^2\|\vh_j\|^2\kappa + 
    \sigma_v^2 
    } 
    \end{split}
\end{equation}
We note that the SIR depends on the number of microphones implicitly through the norm of the ATFs $\|\vh_j\|$, and $\|\vh_0\|$.

Since $\kappa,\rho\ge 0$, and $\kappa > \rho$, the smaller $\mu_j^2,\mu_l^2\;\forall l$ are, the higher the SIR is. 
Using Lemma \ref*{lemma: Riemannian Euclidean mean}, we identify the coefficients for the Riemannian mean as:
\begin{equation}
\begin{split}
\mu_l^{2} =
        \frac{(\sigma_l^2\Vert \vh_l\Vert^2+\sigma_v^2)^{\tau_l}(\sigma_v^2)^{1-\tau_l}-\sigma_v^2}{\Vert \vh_l\Vert^2} , 
\end{split}
\end{equation}
and for the Euclidean mean as
\begin{equation*}
 \mu_l^\text{2} =
 \sigma_l^2\tau_l.
\end{equation*}

It is left to show that 
the coefficients for the Riemannian SIR are smaller than for the Euclidean SIR.
So, we examine the conditions under which
\begin{equation}
\label{eq: mu_j Riem < mu_j Euc explicit expression}
\begin{split}
    \frac{(\sigma_l^2\Vert \vh_l\Vert^2+\sigma_v^2)^{\tau_l}(\sigma_v^2)^{1-\tau_l}-\sigma_v^2}{\Vert \vh_l\Vert^2}    & <
    \sigma_l^2\tau_j \\
    (\sigma_l^2\Vert \vh_l\Vert^2+\sigma_v^2)^{\tau_l}(\sigma_v^2)^{1-\tau_l}    &<
    \sigma_l^2\Vert \vh_l\Vert^2\tau_j + \sigma_v^2 \\
\end{split}
\end{equation}

Equation (\ref{eq: mu_j Riem < mu_j Euc explicit expression}) holds due to the weighted arithmetic mean and geometric mean inequality \cite[pp. 111–112, Theorem 10.5]{cvetkovski2012inequalities} \cite{li2018weighted}.

We see that the Riemannian mean of the correlation matrices leads to the geometric mean of the noise power and the interference power plus the noise power. In contrast, the common practice that is the Euclidean mean of the correlation matrices leads to the arithmetic mean of the powers.

\end{proof}

\subsection{Proof of Lemma \ref{lemma: common eigenvalue equality}}
\begin{proof}
For the Euclidean geometry, the proof is straightforward.

For the Riemannian geometry, since, in general, there is no close form expression for the Riemannian mean, we use its definition as the minimizer of  
\begin{equation}
    \mathbf{\Gamma}_{\text{R}} = \argmin_\mathbf{\Gamma} \sum_j d_{\text{R}}^2(\mathbf{\Gamma},\mathbf{\Gamma}_j).
\end{equation}
For the Affine Invariant metric, the following holds
\begin{equation}
    d_{\text{R}}^2(\mathbf{\Gamma}_1,\mathbf{\Gamma}_2) = 
    \sum_l \log^2 \left(\lambda_l(\mathbf{\Gamma}_1^{-\frac{1}{2}}\mathbf{\Gamma}_2\mathbf{\Gamma}_1^{-\frac{1}{2}})\right).
\end{equation}
Since the matrix $\mathbf{\Gamma}_j^{-\frac{1}{2}}\mathbf{\Gamma}_{\text{R}}\mathbf{\Gamma}_j^{-\frac{1}{2}}$ is a function of the matrices $\mathbf{\Gamma}_j$, we have that $\vh_0$ is also its eigenvector. 
We notice that
\begin{equation}
    \begin{split}
        \min\log^2\left(\lambda_0(\mathbf{\Gamma}_j^{-\frac{1}{2}}\mathbf{\Gamma}_{\text{R}}\mathbf{\Gamma}_j^{-\frac{1}{2}})\right) =
        \min\log^2\left(\frac{\lambda_0(\mathbf{\Gamma}_{\text{R}})}{\lambda_0(\mathbf{\Gamma}_j)}\right) = 0,
    \end{split}
\end{equation}
for $\lambda_0(\mathbf{\Gamma}_{\text{R}}) = \lambda_0(\mathbf{\Gamma}_j)$.
This is the minimum possible value for every $j$, so it must hold that $\lambda_0(\mathbf{\Gamma}_{\text{R}}) = \lambda_0(\mathbf{\Gamma}_j)$.

For the Log-Euclidean distance, we use eigenvalue decomposition and denote by $\{\vu_i\}$ and $\{\vv_k\}$ the eigenvectors of $\mathbf{\Gamma}_{\text{LE}}$ and $\mathbf{\Gamma}_j$, respectively. We have
\begin{equation}
\begin{split}
    d_{\text{LE}}^2(\mathbf{\Gamma}_{\text{LE}},\mathbf{\Gamma}_j) 
    &= 
    \|\ln (\mathbf{\Gamma}_{\text{LE}})-\ln(\mathbf{\Gamma}_j) \|_{\text{F}}^2 \\
    & = 
    \|\sum_i\ln\lambda_i(\mathbf{\Gamma}_{\text{LE}})\vu_i\vu_i^H +
    \ln\lambda_0(\mathbf{\Gamma}_{\text{LE}})\vh_0\vh_0^H \\
    & - 
    \sum_k\ln\lambda_k(\mathbf{\Gamma}_j)\vv_k\vv_k^H  -
    \ln\lambda_0(\mathbf{\Gamma}_{j})\vh_0\vh_0^H \|_{\text{F}}^2  \\
    & =
    \|\mC + 
    (\ln\lambda_0(\mathbf{\Gamma}_{\text{LE}}) - \ln\lambda_0(\mathbf{\Gamma}_{j}))\vh_0\vh_0^H 
    \|_{\text{F}}^2  \\
    &=
    \text{tr}\left(\mC\mC^H \right. \\
    &+ (\ln\lambda_0(\mathbf{\Gamma}_{\text{LE}})  
    -  \left. \ln\lambda_0(\mathbf{\Gamma}_{j}))^2\cdot\|\vh_0\|^2\cdot\vh_0\vh_0^H \right),
\end{split}
\end{equation}
which is minimal for every $j$, for $\lambda_0(\mathbf{\Gamma}_{\text{LE}}) = \lambda_0(\mathbf{\Gamma}_j)$. The last equality is due to the orthogonality between $\vh_0$ and the rest of the eigenvectors.
\end{proof}

\subsection{Proof of Proposition \ref*{prop: derivatives of Riemannian SIR}}
\label{Appendix: Proposition 2, proof and more details}
We start by proving Proposition \ref*{prop: derivatives of Riemannian SIR}, i.e. showing that 
  \begin{equation} 
 \frac{\partial}{\partial\sigma_v^2} \text{SIR}_j(\mathbf{\Gamma}_{\text{R}})   < 
 \frac{\partial}{\partial\sigma_v^2} \text{SIR}_j(\mathbf{\Gamma}_{\text{E}}) < 0,
\end{equation}
and continue with examining the conditions under which
\begin{align}
    \frac{\partial}{\partial\sigma_v^2} \text{SIR}_j(\mathbf{\Gamma}_{\text{R}}) & < 0, \\
    \frac{\partial}{\partial\sigma_v^2} \text{SIR}_j(\mathbf{\Gamma}_{\text{E}}) & < 0.
\end{align}

\begin{proof}
The proof employs the following technical Lemma
\setcounter{lemma}{2}
\begin{lemma}
\label{lemma: c_l>0 and c_l'<0}
For $\mu_j^2$ given by (\ref*{eq: mu_j Riem}), the following holds:
\begin{enumerate}
    \item $\frac{\partial}{\partial\sigma_v^2}\mu_j^{2} \ge 0$ and $\mu_j^{2} \ge 0$.
    \item $\frac{\partial}{\partial\sigma_v^2}\sum_{l=1,l\neq j}^{N_{\text{I}}} 
        \mu_l^2 \|\vh_l\|^2 \ge 0$ and $\sum_{l=1,l\neq j}^{N_{\text{I}}} 
        \mu_l^2 \|\vh_l\|^2 \ge 0$
\end{enumerate}
\begin{proof}
\begin{align}
        \frac{\partial}{\partial\sigma_v^2}\mu_j^2 &= 
        \frac{\partial}{\partial\sigma_v^2}\left( \frac{(\sigma_j^2\Vert \vh_j\Vert^2 + \sigma_v^2)^{\tau_j}(\sigma_v^2)^{1-\tau_j}-\sigma_v^2}{\Vert \vh_j\Vert^2}\right)  \\ &= 
        \frac{1}{\Vert \vh_j\Vert^2} \frac{\partial}{\partial\sigma_v^2}\left((\sigma_j^2\Vert \vh_j\Vert^2 + \sigma_v^2)^{\tau_j}(\sigma_v^2)^{1-\tau_j}-\sigma_v^2 \right) \nonumber \\ &=
        \frac{1}{\Vert \vh_j\Vert^2} \left( {\tau_j}\cdot\left(
        \sigma_j^2\Vert \vh_j\Vert^2(\sigma_v^2)^{\frac{1}{\tau_j}-1}+(\sigma_v^2)^{\frac{1}{\tau_j}}
        \right)^{\tau_j-1}  \right. \nonumber \\ &\cdot
      \left.   \left(\left(\frac{1}{\tau_j}-1\right)\sigma_j^2\Vert \vh_j\Vert^2(\sigma_v^2)^{\frac{1}{\tau_j}-2} + \frac{1}{\tau_j}(\sigma_v^2)^{\frac{1}{\tau_j}-1}\right)
        - 1 
        \right). \nonumber
\end{align}
We focus on the expression in the bracket, rewrite it as a fraction, and show that it is larger than $1$.
\begin{equation}
    \begin{split}
        &\frac{
        \left(\frac{1}{\tau_j}-1\right)\sigma_j^2\Vert \vh_j\Vert^2(\sigma_v^2)^{\frac{1}{\tau_j}-2} + \frac{1}{\tau_j}(\sigma_v^2)^{\frac{1}{\tau_j}-1}
        }
        {
        {\frac{1}{\tau_j}}\cdot\left(
        \sigma_j^2\Vert \vh_j\Vert^2(\sigma_v^2)^{\frac{1}{\tau_j}-1}+(\sigma_v^2)^{\frac{1}{\tau_j}}
        \right)^{1-\tau_j}
        } \\
        &=
        \frac{
        \left(1-\tau_j\right)\sigma_j^2\Vert \vh_j\Vert^2(\sigma_v^2)^{\frac{1}{\tau_j}-2} + (\sigma_v^2)^{\frac{1}{\tau_j}-1}
        }
        {
        \left(
        \sigma_j^2\Vert \vh_j\Vert^2(\sigma_v^2)^{\frac{1}{\tau_j}-1}+(\sigma_v^2)^{\frac{1}{\tau_j}}
        \right)^{1-\tau_j}
        } \\ &=
         \frac{
        \left(1-\tau_j\right)\sigma_j^2\Vert \vh_j\Vert^2(\sigma_v^2)^{\frac{1}{\tau_j}-2} + (\sigma_v^2)^{\frac{1}{\tau_j}-1}
        }
        {
        (\sigma_v^2)^{\frac{1}{\tau_j}-1}\left(
        \sigma_j^2\Vert \vh_j\Vert^2(\sigma_v^2)^{-1}+1
        \right)^{1-\tau_j}
        } \\ &=
        \frac{
        \left(1-\tau_j\right)\sigma_j^2\Vert \vh_j\Vert^2(\sigma_v^2)^{-1} + 1
        }
        {
        \left(
        \sigma_j^2\Vert \vh_j\Vert^2(\sigma_v^2)^{-1}+1
        \right)^{1-\tau_j}
        } \\ &=
        \frac{
        1 + \left(1-\tau_j\right)\sigma_j^2\Vert \vh_j\Vert^2(\sigma_v^2)^{-1}
        }
        {
        \left(
        1 + \sigma_j^2\Vert \vh_j\Vert^2(\sigma_v^2)^{-1}
        \right)^{1-\tau_j}
        } \ge 
        1.
    \end{split}
\end{equation}
The last inequality follows from Bernoullie''s inequality $(1+x)^r \le 1 + rx$, by setting $0 < r = \left(1-\tau_j\right) < 1$, and $x = \sigma_j^2\Vert \vh_j\Vert^2(\sigma_v^2)^{-1} > -1$.
Since $\mu_j^{2}=0$ for $\sigma_v^2=0$ and $\frac{\partial}{\partial\sigma_v^2}\mu_j^{2} \ge 0$, it holds that $\mu_j^{2} \ge 0$.
The above holds for all $j$ thus it also holds for their sum. 
\end{proof}
\end{lemma}

We recall the expression for the SIR from (\ref{eq: SIR expression from proof prop 1}):
\begin{equation}
\begin{split}
    &\text{SIR}_j(\mathbf{\Gamma})  \\
    &=
    1   
    +
    \frac{    
    (\sigma_0^2\|\vh_0\|^2-\mu_j^2\|\vh_j\|^2)\kappa
    + 
        (\mu_j^2\|\vh_j\|^2-\sigma_0^2\|\vh_0\|^2) \rho 
    }
    {
    \sigma_0^2\|\vh_0\|^2\rho + 
        \sum_{l=1,l\neq j}^{N_{\text{I}}} \mu_l^2 \|\vh_l\|^2\rho
        + \mu_j^2\|\vh_j\|^2\kappa + 
    \sigma_v^2  
    } .
\end{split}
\end{equation}
We denote the following functions, which are the numerator and the denominator of the expression of the Riemannian SIR:
\begin{equation}
    f_R(\sigma_v^2) =
         (\sigma_0^2\|\vh_0\|^2-\mu_j^2\|\vh_j\|^2)\kappa  +  
        (\mu_j^2\|\vh_j\|^2-\sigma_0^2\|\vh_0\|^2) \rho 
\end{equation}
\begin{equation}
    g_R(\sigma_v^2) =
      \sigma_0^2\|\vh_0\|^2\rho   + 
        \sum_{l=1,l\neq j}^{N_{\text{I}}} \mu_l^2 \|\vh_l\|^2\rho + \mu_j^2\|\vh_j\|^2\kappa + 
    \sigma_v^2,
\end{equation}
where $\mu_j^2$ is given by (\ref*{eq: mu_j Riem}). Similarly, we define $f_E$, and $g_E$ with $\mu_j^2$ given by (\ref*{eq: mu_j Euc}).
The derivative is
\begin{equation}
    \begin{split}
        \frac{\partial}{\partial\sigma_v^2} \text{SIR}_j(\sigma_v^2) =     
        \frac{f'(\sigma_v^2)g(\sigma_v^2)-g'(\sigma_v^2)f(\sigma_v^2)}{g^2(\sigma_v^2)},
    \end{split}
\end{equation}
where $f$ and $g$ represent $f_R$ or $f_E$ and $g_R$ or $g_E$, respectively, and $(\cdot)'$ denotes the derivative with respect to $\sigma_v^2$.
In the proof of Proposition \ref*{prop: SIR and comparison}, we show that $\mu_j^2$ for the Riemannian mean in (\ref*{eq: mu_j Riem}) is smaller than its Euclidean counterpart in (\ref*{eq: mu_j Euc}). In combination with Lemma \ref{lemma: c_l>0 and c_l'<0}, we have that , $0< g_R(\sigma_v^2)<g_E(\sigma_v^2)$, so we focus on the numerator.

We show that $ \frac{\partial}{\partial\sigma_v^2} \text{SIR}_j(\mathbf{\Gamma}_{\text{R}})    < 
 \frac{\partial}{\partial\sigma_v^2} \text{SIR}_j(\mathbf{\Gamma}_{\text{E}}) < 0 $, by proving the following claims:
 \begin{enumerate}
     \item $f'_R(\sigma_v^2)g_R(\sigma_v^2) < f'_E(\sigma_v^2)g_E(\sigma_v^2) = 0$
     \item $g'_R(\sigma_v^2)f_R(\sigma_v^2) > g'_E(\sigma_v^2)f_E(\sigma_v^2) > 0$
 \end{enumerate}

 \textbf{Proof of Claim 1}
 
 Since $\mu_j^2$ for the Euclidean mean does not depend on $\sigma_v^2$, the same holds for $f_E(\sigma_v^2)$ and $f'_E(\sigma_v^2) = 0$. For the Riemannian case, using Lemma \ref{lemma: c_l>0 and c_l'<0} we have that $f'_R(\sigma_v^2) = \mu_j^{2}\|\vh_j\|^2(\rho - \kappa) < 0$ and $g_R(\sigma_v^2)>0$, so $f'_R(\sigma_v^2)\cdot g_R(\sigma_v^2) < f'_E(\sigma_v^2)\cdot g_E(\sigma_v^2) = 0$.

\textbf{Proof of Claim 2}

Since $\mu_j^2$ for the Riemannian mean in (\ref*{eq: mu_j Riem}) is smaller than its Euclidean counterpart in (\ref*{eq: mu_j Euc}), we have $f_R(\sigma_v^2) > f_E(\sigma_v^2)$.
Additionally, if (\ref*{eq: condition sigma_d>c_l^R}) holds for $\mu_j^2$ in (\ref*{eq: mu_j Euc}) then $f_E(\sigma_v^2) > 0$.

It is left to show that $g'_R(\sigma_v^2)  \ge g'_E(\sigma_v^2)>0$. This holds due to Lemma \ref{lemma: c_l>0 and c_l'<0}:
\begin{equation}
    \begin{split}
         g'_R(\sigma_v^2) 
         &=
        \rho\sum_{l=1,l\neq j}^{N_{\text{I}}} 
        \frac{\partial}{\partial\sigma_v^2}\mu_l^2 \|\vh_l\|^2 +
        \|\vh_j\|^2\kappa\frac{\partial}{\partial\sigma_v^2}\mu_j^{2} +
        1 \\ 
        & \ge 
        1
         =  g'_E(\sigma_v^2).
    \end{split}
\end{equation}

\end{proof}
Following the proof of Proposition \ref*{prop: derivatives of Riemannian SIR}, since $f'_E(\sigma_v^2)=0$, it follows that iff $g'_E(\sigma_v^2)f_E(\sigma_v^2) > 0$ then $\frac{\partial}{\partial\sigma_v^2} \text{SIR}_j(\mathbf{\Gamma}_{\text{E}})  < 0$.
Since $g'_E(\sigma_v^2)=1$, we have that
$\frac{\partial}{\partial\sigma_v^2} \text{SIR}_j(\mathbf{\Gamma}_{\text{E}})  < 0$ iff $f_E(\sigma_v^2) > 0$. From the expression of $f_E(\sigma_v^2)$ it follows that 
$f_E(\sigma_v^2) > 0$ iff condition (\ref*{eq: condition sigma_d>c_l^R}) holds. So, it is a sufficient and a necessary condition for $\frac{\partial}{\partial\sigma_v^2} \text{SIR}_j(\mathbf{\Gamma}_{\text{E}})  < 0$ to hold. 

For the Riemannian mean, the condition under which $\frac{\partial}{\partial\sigma_v^2} \text{SIR}_j(\mathbf{\Gamma}_{\text{R}})  < 0$ is more easily met. Additionally, it is a sufficient but not a necessary condition, as we show next.

First, we note that condition (\ref*{eq: condition sigma_d>c_l^R}) can be recast as 
\begin{equation}
\label{eq: sigma h > mu h}
 \begin{split}
     \sigma_0^2\|\vh_0\|^2 \ge \mu_j\|h_j\|^2, \;\; \forall j,
 \end{split}
 \end{equation}
where $\mu_j^2$ given by (\ref*{eq: mu_j Euc}). 

For the Riemannian mean, under condition (\ref{eq: sigma h > mu h}) with $\mu_j^2$ given by (\ref*{eq: mu_j Riem}) it holds that $g'_R(\sigma_v^2)f_R(\sigma_v^2) > 0$. Since $f'_R(\sigma_v^2)g_R(\sigma_v^2) < 0$,
condition (\ref{eq: sigma h > mu h}) with $\mu_j^2$ given by (\ref*{eq: mu_j Riem}) is a sufficient but not a necessary condition for $\frac{\partial}{\partial\sigma_v^2} \text{SIR}_j(\mathbf{\Gamma}_{\text{R}})  < 0$ to hold, as opposed to the Euclidean case.  

In the proof of Proposition \ref*{prop: SIR and comparison} it is shown that the parameter $\mu_j$ for the Riemannian mean in (\ref*{eq: mu_j Riem}) is smaller than its Euclidean counterpart in (\ref*{eq: mu_j Euc}). Therefore, condition (\ref{eq: sigma h > mu h}) is more easily met when $\mu_j$ is given by (\ref*{eq: mu_j Riem}) than when $\mu_j$ is given by (\ref*{eq: mu_j Euc}). As a consequence, it is possible for $\frac{\partial}{\partial\sigma_v^2} \text{SIR}_j(\mathbf{\Gamma}_{\text{E}})$ to be positive, while $\frac{\partial}{\partial\sigma_v^2} \text{SIR}_j(\mathbf{\Gamma}_{\text{R}})$ is negative.

\subsection{Proof of Proposition \ref*{prop: general SIR  Gamma_R superiority}}
 \begin{proof}
Under Assumption \ref*{assump: uncorrelated ATFs of target and interferences} the ATF of the desired source, $\vh_0$, is an eigenvector of $\mathbf{\Gamma}_j$ for all $j$ with the same eigenvalue $\lambda_0$. Then, according to Lemma \ref*{lemma: common eigenvalue equality} the following holds
\begin{equation}
\begin{split}
    \vh_0^H\mathbf{\Gamma}_{\text{R}}\vh_0 = 
    \vh_0^H\mathbf{\Gamma}_{\text{E}}\vh_0=\lambda_0 .
\end{split}
\end{equation} 
For the Riemannian and the Euclidean mean, the following holds \cite{lim2012matrix}:
\begin{equation}
    \mathbf{\Gamma}_{\text{R}} \preceq \mathbf{\Gamma}_{\text{E}}.
\end{equation}
Thus, for all $j$:
\begin{equation}
    \vh_j^H\mathbf{\Gamma}_{\text{R}}\vh_j \le 
    \vh_j^H\mathbf{\Gamma}_{\text{E}}\vh_j
\end{equation}
and therefore
\begin{equation}\sum_{j=1}^{N_\text{I}}\vh_j^H\mathbf{\Gamma}_{\text{R}}\vh_j \le
    \sum_{j=1}^{N_\text{I}}\vh_j^H\mathbf{\Gamma}_{\text{E}}\vh_j,
\end{equation}
which completes the proof. 

   \end{proof}

\subsection{Proof of Proposition \ref*{prop: optimal offset for partition}}
\begin{proof}
The Riemannian mean is $\mathbf{\Gamma}_{\text{R}} = \mathbf{\Gamma}_1^{\frac{1}{2}}\mathbf{\Gamma}_2^{\frac{1}{2}}$, since the matrices commute. The matrices $\mathbf{\Gamma}_1$ and $\mathbf{\Gamma}_2$ are expressed using their eigenvectors, allowing the derivation of $\mathbf{\Gamma}_1^{\frac{1}{2}}$ and $\mathbf{\Gamma}_2^{\frac{1}{2}}$. Finally, $\mathbf{\Gamma}_{\text{R}}$ is computed via their product:
\begin{equation}
    \begin{split}
        \mathbf{\Gamma}_{\text{R}} = 
        \sigma_d^2\vh^d(\vh^d)^H + 
        \mu_1^2 \vh^i_1(\vh^i_1)^H + 
         \mu_2^2
        \vh^i_2(\vh^i_2)^H + 
        \sigma_v^2 \mI
    \end{split}
\end{equation}
with $\mu_j^2 = \frac{((\alpha^2\sigma_j^2\|\vh_j\|^2+\sigma_v^2)^{\frac{1}{2}} \cdot ((1-\alpha)^2\sigma_j^2\|\vh_j\|^2+\sigma_v^2)^{\frac{1}{2}})-\sigma_v^2 }{\|\vh_j\|^2}$ for $j=1,2$.

For the Euclidean mean, $\mu_j^2 =  \frac{\sigma_j^2}{2} (\alpha^2+(1-\alpha)^2)$, which reaches its minimum for $\alpha=\frac{1}{2} \; \forall j$.

From the proof of Proposition \ref*{prop: SIR and comparison}, the smaller the $\mu_j^2$-s are, the higher the SIR is. 
We show that
\begin{equation}
\begin{split}        &       \frac{((\alpha^2\sigma_j^2\|\vh_j\|^2+\sigma_v^2)^{\frac{1}{2}} \cdot ((1-\alpha)^2\sigma_j^2\|\vh_j\|^2+\sigma_v^2)^{\frac{1}{2}})-\sigma_v^2 }{\|\vh_j\|^2} \\ 
&\le 
\frac{\sigma_j^2}{2} (\alpha^2+(1-\alpha)^2). 
 \end{split}
\end{equation}
Rearranging results in
\begin{equation}
    \begin{split}
        &((\alpha^2\sigma_j^2\|\vh_j\|^2+\sigma_v^2)^{\frac{1}{2}} \cdot ((1-\alpha)^2\sigma_j^2\|\vh_j\|^2+\sigma_v^2)^{\frac{1}{2}}) \\ &\le 
        \|\vh_j\|^2\frac{\sigma_j^2}{2} (\alpha^2+(1-\alpha)^2) +
        \sigma_v^2,
    \end{split}
\end{equation}
which holds due to the inequality of arithmetic and geometric means $\sqrt{x \cdot y} \le \frac{x+y}{2}$, where $x = \alpha^2\sigma_j^2\|\vh_j\|^2+\sigma_v^2$ and $y = (1-\alpha)^2\sigma_j^2\|\vh_j\|^2+\sigma_v^2$.

\end{proof}

\section*{Theoretical Results For the Total SIR}
\label{Appendix: results for total SIR}
\begin{proposition}
\label{prop: SIR and comparison for total sir}
For any number of microphones in the array, the following holds
\begin{equation}
    \text{SIR}_{\text{tot}}(\mathbf{\Gamma}_{\text{R}}) > \text{SIR}_{\text{tot}}(\mathbf{\Gamma}_{\text{E}}).
\end{equation}

\begin{proof}
We compute the output total SIR for a general correlation matrix with a similar structure as in Lemma \ref*{lemma: Riemannian Euclidean mean}:
\begin{equation}
    \begin{split}
        \mathbf{\Gamma} = 
        \sigma_0^2\vh_0^H\vh_0 + 
        \sum_{l=1}^{N_{\text{I}}} \mu_l^2 \vh_l^H\vh_l +
    \sigma_v^2\mI,
    \end{split}
\end{equation}
and the SIR becomes:
%
\begin{equation}
    \begin{split}
        \text{SIR}_{\text{tot}}(\mathbf{\Gamma})  
        & =
        \frac{
        \vd_0^H\left(
        \sigma_0^2\vh_0^H\vh_0 + 
        \sum_{l=1}^{N_{\text{I}}} \mu_l^2 \vh_l^H\vh_l +
    \sigma_v^2\mI
    \right)\vd_0
    }
    { \frac{1}{N_{\text{I}}}\sum_{j=1}^{N_{\text{I}}} 
     \vd_j^H\left(
        \sigma_0^2\vh_0^H\vh_0 + 
        \sum_{l=1}^{N_{\text{I}}} \mu_l^2 \vh_l^H\vh_l +
    \sigma_v^2\mI
    \right)\vd_j 
    }                          \\
    &=
    \frac{
    \sigma_0^2|\langle\vd_0^H,\vh_0\rangle|^2 + 
        \sum_{l=1}^{N_{\text{I}}} \mu_l^2 |\langle\vd_0^H,\vh_l\rangle|^2 +
    \sigma_v^2 M 
    }
    {  \frac{1}{N_{\text{I}}}\sum_{j=1}^{N_{\text{I}}} \left(
    \sigma_0^2|\langle\vd_j^H,\vh_0\rangle|^2 + 
        \sum_{l=1}^{N_{\text{I}}} \mu_l^2 |\langle\vd_j^H,\vh_l\rangle|^2 +
    \sigma_v^2 M
    \right)
    }                  \\
   & =
  \frac{
  \sigma_0^2M\|\vh_0\|^2\kappa+ 
        \sum_{l=1}^{N_{\text{I}}} \mu_l^2 M \|\vh_l\|^2 \rho +
    \sigma_v^2 M 
    }
    {     {
    \sigma_0^2 M \|\vh_0\|^2\rho + 
     \frac{1}{N_{\text{I}}}\sum_{j=1}^{N_{\text{I}}}
        \sum_{l=1,l\neq j}^{N_{\text{I}}} \mu_l^2 M \|\vh_l\|^2\rho } {
        + \frac{1}{N_{\text{I}}}\sum_{j=1}^{N_{\text{I}}}
        \mu_j^2 M \|\vh_j\|^2\kappa + 
    \sigma_v^2 M }
    }                      \\
     & =
    1  +
    \frac{ {
    (\sigma_0^2\|\vh_0\|^2 - \frac{1}{N_{\text{I}}}\sum_{j=1}^{N_{\text{I}}}
    \mu_j^2\|\vh_j\|^2)\kappa+ 
        (\frac{1}{N_{\text{I}}}\sum_{j=1}^{N_{\text{I}}}
        \mu_j^2\|\vh_j\|^2 }{ -\sigma_0^2\|\vh_0\|^2) \rho }
    }
    {{
    \sigma_0^2\|\vh_0\|^2\rho + 
        \frac{N_{\text{I}}-1}{N_{\text{I}}}\sum_{l=1}^{N_{\text{I}}} \mu_l^2 \|\vh_l\|^2\rho +
        \frac{1}{N_{\text{I}}}\sum_{j=1}^{N_{\text{I}}} 
        \mu_j^2\|\vh_j\|^2\kappa } { + 
    \sigma_v^2 }
    } .
    \end{split}
\end{equation}

In the proof of Proposition \ref*{prop: SIR and comparison} we show that $\mu_j^2$ for the Riemannian mean is smaller than $\mu_j^2$ for the Euclidean mean for all $j$. Consequently, we have that $\sum_{j=1}^{N_{\text{I}}}
        \mu_j^2\|\vh_j\|^2$ for the Riemannian mean is smaller than for the Euclidean mean. 
Since $\kappa > \rho \ge 0$, it holds that $\text{SIR}_{\text{tot}}(\mathbf{\Gamma}_{\text{R}}) > \text{SIR}_{\text{tot}}(\mathbf{\Gamma}_{\text{E}}) $.

\end{proof}

\end{proposition}

\begin{proposition}
If 
\begin{equation}
 \begin{split}
     \sigma_0^2\|\vh_0\|^2 \ge
     \frac{1}{N_{\text{I}}}\sum_{j=1}^{N_{\text{I}}}
     \sigma_j^{2}\tau_j\|h_j\|^2,
 \end{split}
 \end{equation}
then
  \begin{equation} 
 \frac{\partial}{\partial\sigma_v^2} \text{SIR}_{\text{tot}}(\mathbf{\Gamma}_{\text{R}})   < 
 \frac{\partial}{\partial\sigma_v^2} \text{SIR}_{\text{tot}}(\mathbf{\Gamma}_{\text{E}}) < 0.
\end{equation}
\begin{proof}
The proof follows the same steps as the proof of Proposition \ref*{prop: derivatives of Riemannian SIR}
\end{proof}
\end{proposition}





\end{document}